\DocumentMetadata{}
\documentclass[sigconf, nonacm]{acmart}

\usepackage{acmart-taps}
\usepackage{xspace}
\usepackage{enumitem}
\usepackage{color, xcolor}
\usepackage{caption}
\usepackage{tabularx}
\usepackage{longtable}
\usepackage{tcolorbox}
\tcbuselibrary{skins}
\usepackage{hyperref}
\usepackage{xurl}

\def\insitu{\textit{in situ}\xspace}
\def\ie{\textit{i.e.,}\xspace}
\def\etal{\textit{et~al.}\xspace}

\def\eg{\textit{e.g.,}\xspace}
\def\incl{\textit{incl.}\xspace}
\def\vs{\textit{vs.}\xspace}

\definecolor{darkblue}{RGB}{0, 51, 153}

\author{Chen Chen}
\orcid{0000-0001-7179-0861}
\email{chechen@fiu.edu}
\affiliation{%
  \institution{Florida International University}
  \city{Miami}
  \state{FL}
  \country{USA}
}

\author{Lingyao Li}
\orcid{0000-0001-5888-8311}
\email{lingyaoli@arizona.edu}
\affiliation{%
  \institution{University of Arizona}
  \city{Tucson}
  \state{AZ}
  \country{USA}
}

\author{Renkai Ma}
\orcid{0000-0002-4434-2235}
\email{mark@ucmail.uc.edu}
\affiliation{%
  \institution{University of Cincinnati}
  \city{Cincinnati}
  \state{OH}
  \country{USA}
}

\author{Rawan Alghofaili}
\orcid{0000-0001-6510-4562}
\email{rawan@utdallas.edu}
\affiliation{%
  \institution{University of Texas at Dallas}
  \city{Richardson}
  \state{TX}
  \country{USA}
}

\author{Shaoze Zhou}
\orcid{0009-0000-3243-0599}
\email{szhou010@fiu.edu}
\affiliation{%
  \institution{Florida International University}
  \city{Miami}
  \state{FL}
  \country{USA}
}

\author{Bojun Zhang}
\orcid{0009-0006-4312-5032}
\email{bzhan035@fiu.edu}
\affiliation{%
  \institution{Florida International University}
  \city{Miami}
  \state{FL}
  \country{USA}
}

\author{Xian Su}
\orcid{0000-0001-5903-6380}
\email{xsu@fiu.edu}
\affiliation{%
  \institution{Florida International University}
  \city{Miami}
  \state{FL}
  \country{USA}
}

\author{Weidong Zhu}
\orcid{0000-0002-9812-6634}
\email{weizhu@fiu.edu}
\affiliation{%
  \institution{Florida International University}
  \city{Miami}
  \state{FL}
  \country{USA}
}

\author{Christine Lisetti}
\orcid{0000-0003-0756-133X}
\email{lisetti@fiu.edu}
\affiliation{%
  \institution{Florida International University}
  \city{Miami}
  \state{FL}
  \country{USA}
}

\author{Mo Sha}
\orcid{0000-0002-2701-0159}
\email{msha@fiu.edu}
\affiliation{%
  \institution{Florida International University}
  \city{Miami}
  \state{FL}
  \country{USA}
}


\begin{teaserfigure}
    \centering
    \includegraphics[width=\linewidth]{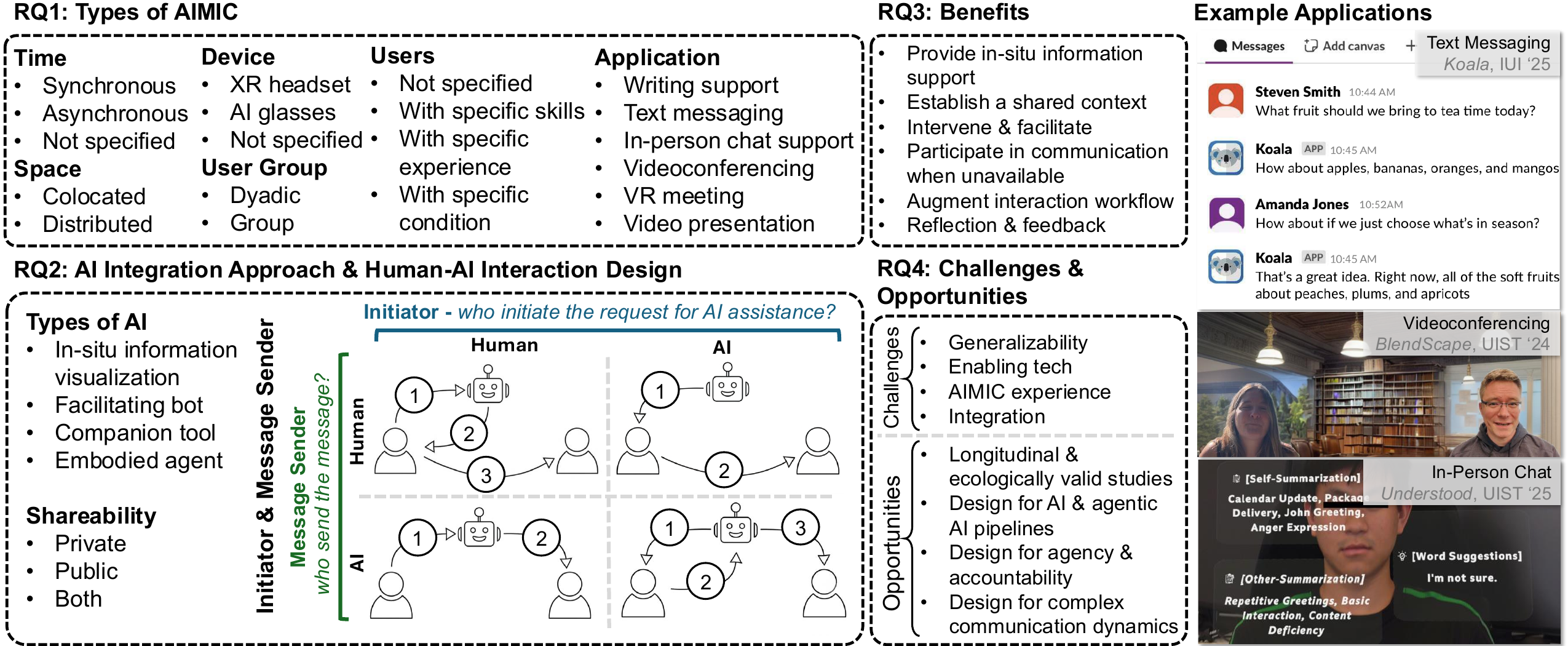}
    \vspace{-0.3in}
    \caption{An overview of the key design dimensions that we examined in AIMIC. Figures of example applications are taken from \cite{Houde2025}, \cite{Rajaram2024BlendScape}, and \cite{Zhang2025Understood}, in top-to-bottom order.}
    \label{fig:teaser}
\end{teaserfigure}

\begin{document}

%TC:ignore

\title[Understanding the Design Taxonomy of AIMIC Experiences in HCI]{Understanding the Design Taxonomy of AI-Mediated Interpersonal Communication Experiences in HCI: A Scoping Analysis}


\begin{abstract}

Interpersonal communication is a fundamental aspect of everyday life, shaping interactions across workplaces, education, entertainment, healthcare, and beyond.
While computer-mediated communication has been extensively studied, a comprehensive understanding of \textbf{AI}-\textbf{M}ediated \textbf{I}nterpersonal \textbf{C}ommunication~(AIMIC) remains lacking.
An in-depth scoping analysis is urgently needed to understand the research landscape of AIMIC in HCI, particularly following the release of ChatGPT, the rapid growth of large foundation models, and AI agent research.
We conducted a comprehensive scoping analysis to understand AIMIC by performing an in-depth review of prior HCI literature published over the past decade (January, 2016 - May, 2026).
Grounded in the \textbf{P}referred \textbf{R}eporting \textbf{I}tems for \textbf{S}ystematic reviews and \textbf{M}eta-\textbf{A}nalyses (PRISMA) approach, we curated $52$~full-paper publications from the HCI literature spanning a range of interpersonal communication contexts.
We analyzed this corpus by examining the types of AIMIC studied, AI integration approaches and human-AI interaction design, reported outcomes and benefits, as well as key challenges and future research opportunities.

\end{abstract}

\renewcommand{\shortauthors}{Chen~\etal}
\pagenumbering{arabic}
\settopmatter{printfolios=true}
\setcopyright{none}

\maketitle

%TC:endignore

\section{Introduction}\label{sec::introduction}

\begin{quote}
    {\it ``The medium is the message.''} -- Marshall McLuhan, {\it Understanding Media} (1964)~\cite{McLuhan1964}
\end{quote}

Far from a mere exchange of information, interpersonal communication is the indispensable currency of the modern world, spanning across workspace, education, entertainment, healthcare and beyond.
Advances in digital tools and the internet have enabled a wide range of forms of \textbf{C}omputer-\textbf{M}ediated \textbf{C}ommunication (CMC).
CMC manifests in various ways, as described in the long-standing \textbf{C}omputer \textbf{S}upported \textbf{C}ooperative \textbf{W}ork (CSCW) matrix, which organizes communication along the dimension of time and space~\cite{Johansen2020, Rodden1991}.
CMC can occur in dyadic settings between two people or within larger groups. 
Furthermore, the communication experience can be in-person or distributed, and take place either synchronously or asynchronously.
Facilitating engaging and effective conversation presents several challenges. 
For example, participants often struggle to connect when there is a significant information asymmetry or a lack of shared background knowledge; as the number of participants grows, ensuring inclusivity becomes increasingly difficult due to the complexities of the `many-mind problem'~\cite{Cooney2020}. 
These challenges can be particularly pronounced for individuals who face verbal communication barriers or experience social withdrawal~\cite{Wiklund2016}.

Recent advances in foundational AI models have opened new avenues for designing and integrating AI across a broad spectrum of CMC, often referred to as \textbf{AI}-\textbf{M}ediated \textbf{C}ommunication~(AIMC).
The integrated AI and AI agents typically leverages pretrained foundation models to pursue goals and execute tasks on behalf of users~\cite{aiagents, Qu2025}. These agents exhibit core capabilities such as reasoning, planning, and memory, while operating with varying levels of autonomy to make decisions, learn, and adapt to new contexts~\cite{aiagents, Qu2025}.
Since the introduction of ChatGPT in November 2022~\cite{chatgptnews}, a growing range of AI capabilities has been integrated into commercially available communication tools, supporting diverse forms of communication across multiple modalities.
For example, videoconferencing tools such as Teams~\cite{TeamsCoPilot} and Zoom~\cite{ZoomAICompanion} have integrated AI as a companion feature that allows meeting participants to query contextual meeting information, obtain additional background details, or conduct post-meeting reflection.
\textbf{I}nstant \textbf{M}essaging~(IM) applications such as WhatsApp have introduced the ``writing help'' feature to assist users in improving their text-based communication~\cite{whatsappwritinghelp}.
While extensive prior work has systematically reviewed the design, systems, and user experiences of CMC~\cite{Metz1994, Rains2016, Nowak2018, Lee2017} as well as human–AI interaction~\cite{Kulkarni2019, Deng2025, Kusal2022, Bhardwaj2024}, our understanding of complex AIMC remains limited, more specifically, how AI can mediate and facilitate interpersonal communications.
While the design of efficient and effective AIMC will draw on a diverse range of human–AI interaction techniques, the primary focus remains on augmenting the interpersonal communication experience.

This paper conducted a comprehensive survey investigating the design of \textbf{AI}-\textbf{M}ediated \textbf{I}nterpersonal \textbf{C}ommunication (AIMIC).
Unlike existing surveys that explore CMC and interactions with AI agents, our focus is situated within a range of interpersonal communication contexts inspired by the long-standing CSCW matrix.
Our analysis also accounts for the number and physical distribution of participants, the nature of tasks, synchronicity, and communication modalities.
AIMIC can be considered a specific case of AIMC, focusing exclusively on interpersonal communication while excluding mass communication contexts.
In contrast to mass communication, which is broad and impersonal (\eg~through social media), interpersonal communication involves direct, one-to-one or small-group exchanges of messages between individuals~\cite{Chaffee1982}.
While mass communication is often included in discussions of CMC, this survey focuses specifically on interpersonal communication~\cite{Sundar2022, Chaffee1982}.
Grounded on the \textbf{P}referred \textbf{R}eporting \textbf{I}tems for \textbf{S}ystematic reviews and \textbf{M}eta-\textbf{A}nalyses (PRISMA) framework~\cite{page2021prisma, page2022prisma}, we conducted a systematic literature review of $52$~publications in the field of \textbf{H}uman-\textbf{C}omputer \textbf{I}nteraction (HCI) from January 2016 to May 2026. This period begins shortly after the release of TensorFlow in late 2015~\cite{aihistory}, which helped democratize the use of AI across research fields, and spans the emergence of ChatGPT in late 2022~\cite{chatgptnews}, which significantly accelerated the growth and visibility of AI applications.
Guided by our goal, we aim to address four Research Questions (RQs):

\vspace{4px}
\begin{itemize}[topsep=0pt, itemsep=0pt, leftmargin=9.5pt]

    \item {\bf RQ1 - Types of AIMIC Studied:} Which forms of AIMC are investigated in the current literature?

    \item {\bf RQ2 - AI Integration Approach and Human-AI Interaction Design:} How has AI been integrated into different forms of CMC, and how can these approaches be systematically organized? 

    \item {\bf RQ3 - Outcomes and Benefits: } What are the outcomes and benefits for the AIMIC experiences explored in current literature?
    
    \item {\bf RQ4 - Challenges and Opportunities:} What key challenges and opportunities that have been identified?

\end{itemize}

Our scoping analysis identifies $13$ dimensions, categorized across the above four RQs.
Figure~\ref{fig:teaser} presents an overview of the resulting dimensions along with representative applications drawn from the curated literature~\cite{Houde2025, Rajaram2024BlendScape, Zhang2025Understood}.
Our findings reveal a research landscape focused on synchronous and distributed communication, with recent work increasingly shifting from traditional AI techniques toward LLM-powered and more agentic forms of mediation. 
We found that AI most commonly provides \insitu~information support or actively intervenes to facilitate communication, while taking on diverse roles in initiating, reformulating, augmenting, and delivering communicative content. 
Our analysis further identifies four recurring challenges and four corresponding opportunities. 
Together, these findings offer a structured foundation for HCI researchers and practitioners to critically examine existing AIMIC systems and inform the design of future ones.
\section{Related Works}\label{sec::related-works}

In this section, we review related work on interpersonal communication and CMC (Section~\ref{sec::related::cmc}) as well as AIMC (Section~\ref{sec::related::aimc}).
While focusing on AIMIC, the analysis of our survey will be grounded in established constructs from interpersonal communication and CMC.

\subsection{Interpersonal Communication and Computer-Mediated Communication~(CMC)}\label{sec::related::cmc}

Tubbs~\cite{Tubbs2012} distinguishes between interpersonal communication, characterized by direct interaction and reciprocal feedback, and mass communication, which is broadcast to large anonymous audiences.
Interpersonal communication is ubiquitous and plays a critical role in everyday life, serving as a fundamental tool for bonding, collaboration, conflict resolution, idea sharing, and inspiration.
Burleson~\cite{Burleson2010} defines the process of interpersonal communication as {\it ``a complex, situated social process in which people who have established a communicative relationship exchange \textbf{messages} in an effort to generate shared meanings and accomplish social goals.''}
Beyond words and sentences, the {\it ``messages''} is fundamentally a speech act, referring to the performance of actions through the expressions of verbal and non-verbal cues~\cite{Tracy2013}.
Despite being central to both our personal and professional lives, interpersonal communication can be challenging.
Successfully transmitting a message from sender to receiver is influenced by context, personal filters (\eg~ beliefs, emotions, and experiences), and feedback~\cite{communication_challenge}. A breakdown at any one of these stages can degrade the quality of the communication experience.
In group conversations, these challenges are often amplified by long-standing ``many-mind problems''~\cite{Cooney2020} that undermine performance, such as bias, fear of speaking up, and unfocused discussion~\cite{Bhattacharya2018}.

Advances in and the democratization of digital devices and the internet have enabled diverse forms of interpersonal communication, giving rise to a new stream of research on CMC~\cite{Yao2020, Liang2015}.
McQuail~\cite{McQuail2010} defines CMC as any act of communication that takes place through the use of two or more electronic devices. 
Despite the very broad definition, the concept of CMC is generally conceptualized as consisting of specific characteristics or affordances that contrast traditional face-to-face interpersonal communication (\eg~email, text messaging, social network site interactions, videoconferencing)~\cite{Liang2015, Thurlow2004}.
The diverse forms of interpersonal communication and CMC can be classified by the long-standing CSCW matrix based on two key dimensions - \emph{time} (synchronous \vs~asynchronous) and \emph{space} (co-located \vs~distributed)~\cite{Johansen2020, Rodden1991}.
While CSCW was originally introduced in the mid-1980s to describe the growing interest in using computer technologies to support group activities~\cite{Grudin1994, Poltrock1994}, the concept has since been widely adopted across a broad range of CMC research.

A number of systematic reviews have examined different forms of CMC. 
An early survey by Metz~\cite{Metz1994} reviewed CMC tools up to the 90s across organizational, interpersonal, and mass communication contexts. 
Rains~\etal~\cite{Rains2016} examined the potential benefits and drawbacks of CMC for social support processes. 
Nowak~\etal~\cite{Nowak2018} provided a scoping review of the design and use of digital representations in CMC, while Tang~\etal~\cite{Tang2019} focused on the use of emoticons, emojis, and stickers. 
Other work has also explored CMC in specific settings, such as patient-doctor communication~\cite{Lee2017} and the design of smart meeting rooms~\cite{Freitas2015, Yu2010SmartMeetingSystems}.
Identifying opportunities and challenges in the transition from CMC to AIMC is not straightforward. While we acknowledge that our curated corpus can be viewed as instantiations of CMC, our focus is on the design and integration of AI into diverse forms of interpersonal communication.

\subsection{AI-Mediated Interpersonal Communication (AIMIC)}\label{sec::related::aimc}

The introduction of AI into interpersonal communication has the potential to positively transform how people interact and communicate. 
Recent friction AI framework also highlights how intentional design can promote reflection and engagement by introducing productive friction, rather than solely prioritizing frictionless automation~\cite{Natali2024a, Natali2024b}.
AIMC is defined as the {\it ``mediated communication between people in which a computational agent operates on behalf of a communicator by \textbf{modifying}, \textbf{augmenting}, or \textbf{generating} messages to accomplish communication or interpersonal goals''}~\cite{Hancock2020}.
Mirroring the framework of CMC, AIMC emphasizes how interpersonal exchanges can be supported and enhanced through various forms of AI integration~\cite{Hancock2020}. 
Unlike traditional CMC, interpersonal communication is no longer merely transmitted through technology; instead, it can be modified, augmented, or even generated by computational agents to achieve communicative goals~\cite{Hancock2020}.
Sundar~\etal~\cite{Sundar2000} distinguished AI-mediated communication from CMC by focusing on ``source orientation;'' for example, when individuals interact with computers, is the machine the source of communication and the object of interaction, or is it simply a medium or channel through which two or more humans communicate?
Building on the concept of AIMC, we use AIMIC to refer specifically to interpersonal communication, excluding mass communication.
Lee~\etal~\cite{Lee2025} identified four distinct AIMIC patterns (Figure~\ref{fig:teaser}): \textbf{(1)} humans can request and relay AI-generated content~\cite{Hancock2020}; \textbf{(2)} humans can selectively share AI-generated insights or viewpoints to their communication partners~\cite{Do2022}; \textbf{(3)} AI can reformulate and present messages provided by humans~\cite{Wang2022, Natarajan2025}; and \textbf{(4)} AI can directly solicit and share input from one or more communication participants~\cite{Wang2022, Natarajan2025}.
Existing constructs have highlighted both the challenges of AIMIC design and the consequences of poorly designed AIMIC systems; for example, Media Richness Theory~\cite{Daft1986} suggests that the high information capacity of face-to-face communication can be disrupted by AI interventions.

While the long-standing CSCW matrix provides a useful foundation for understanding CMC, AIMIC may require consideration of additional dimensions.
Hancock~\etal~\cite{Hancock2020} propose six dimensions for characterizing the design of AIMC systems, including \emph{magnitude}, \emph{media type}, \emph{optimization goal}, \emph{autonomy}, and \emph{role orientation}; however, the primary work it builds on is primarily focused on distributed, asynchronous communication (\eg~email and text messaging).
We argue that synchronous interpersonal communication - whether co-located (\eg~ augmented by Mixed Reality headset \cite{Zhou2026ChatMuseEA, Zhou2026ChatMuse}, AI Glasses~\cite{Yang2025SocialMind} or shared ambient display) or distributed (\eg~videoconferencing) - should also be recognized as a central domain.
Arets~\etal~\cite{Arets2025} present a scoping review on the role of generative AI in facilitating social interaction; however, their survey primarily focuses on generative AI-mediated synchronous and distributed communication experiences, leaving other forms of interpersonal communication and the use of traditional AI approaches, such as topic modelling and key entity extraction largely unexamined.
Sundar~\etal~\cite{Sundar2022} classify AI's involvement in interpersonal communication by examining both mass and interpersonal communication contexts.
Other studies have examined affect-related impacts in AIMIC, such as perceptions of authenticity, sincerity, and trust~\cite{Hohenstein2020, Jakesch2019, Sahebi2025}; factors that may diminish communication quality~\cite{Hohenstein2023, Tafazoli2024}; and AIMIC in specific application domains like online learning~\cite{Wang2022}.

Through a systematic in-depth literature survey, this work aims to understand how AI can be designed and integrated into various forms of interpersonal communications.
While mass communication is often included in discussions of CMC, we focus specifically on interpersonal communication~\cite{Sundar2022}.
Much prior research has identified the challenges in designing AIMIC. 
For example, the paradigm of situated interaction has identified several key challenges, including modeling interaction initiatives, contextual interpretation, grounding, and turn-taking~\cite{Dey2001SituatedInteraction, Bohus2009}.
Our scope differs from existing scoping reviews on human-AI communication and interaction~\cite{Kulkarni2019, Deng2025, Kusal2022, Bhardwaj2024} in that we focus specifically on interpersonal communication. 
However, we acknowledge that some projects in our curated corpus may introduce innovative approaches to human–AI interaction.

\section{Method}\label{sec::method}

To systematically understand how AI can be designed and integrated into the workflow and experience of interpersonal communications, we conducted a systematic literature review of publications over the past decade (January, 2016 - May, 2026).
This was around the time when Google released TensorFlow in late 2015 - one of the most widely adopted frameworks that helped democratize AI and broaden its accessibility across diverse research fields~\cite{aihistory}.

\subsection{Data Collection}\label{sec::method::datacollection}

Our data collection process followed a multi-stage search and screening protocol inspired by the PRISMA methodology~\cite{page2021prisma, page2022prisma} to construct a comprehensive and relevant corpus of papers. 
While focusing on AI-mediated interpersonal communication, the specific communication scenarios can be highly heterogeneous.
The term ``AI-mediated interpersonal communication'' may not uniformly used in the literature that matches our focus.
Consequently, we had to build a broader query around a range of adjacent and relevant terms.
Table~\ref{tab::keywords} presents three sets of keywords that were iteratively developed over six months to capture different aspects of AI-mediated communication support systems within three HCI and AI research groups.
We treat each term and its commonly used abbreviation, such as ``Mixed Reality'' and ``MR'' as distinct entries.
Our final query is composed of these five sets and was constructed as follows:

% \begin{center}
%     (\texttt{``AI-mediated communication'' ||} \\ \texttt{(Set A \&\& Set B \&\&  Set C})).
% \end{center}

\begin{center}
    {\fontsize{8.63}{12.5}\selectfont (\texttt{``AI-mediated communication'' || (Set A \&\& Set B \&\& Set C)})}
\end{center}

Only the full-paper peer-reviewed publications over the past decade (January, 2016 - May, 2026) are considered, including journal articles and conference full papers. 
Extended abstracts, work-in-progress papers, and non-peer-reviewed preprint were excluded, as they typically present early-stage concepts or preliminary ideas without comprehensive evaluation.
We focus exclusively on papers published through IEEE Xplore\footnote{IEEE Xplore Digital Library: \href{https://ieeexplore.ieee.org/Xplore/home.jsp}{https://ieeexplore.ieee.org/Xplore/home.jsp}. Accessed on May, 16, 2026.}, ACM Digital Library\footnote{ACM Digital Library: \href{https://dl.acm.org}{https://dl.acm.org}. Accessed on May 16, 2026.}
, and Taylor \& Francis Group\footnote{Taylor \& Francis Group: \href{https://www.taylorfrancis.com}{https://www.taylorfrancis.com}. Accessed on May 16, 2026.}
, which are major publishers of high-quality research in HCI, Extended Reality, Visualization, and Accessibility.
While domain-specific applications such as healthcare an important area of AIMIC, we only focus on AIMIC experience generalizable across broader context, and therefore excludes domain-specific venues from its scope.
Our data collection was conducted on June 10, 2026.

\begin{table*}[htbp]
    \centering
    \caption{Search queries and keywords.}
    \vspace{+4px}
    \begin{tabularx}{\textwidth}{llX}
        \toprule
         & \textbf{Dimension} & \textbf{Key Words} \\ 
        \midrule
        \textbf{Set A} & Task & conversation, chat, collaboration, communication, meeting, videoconferencing, training, instruction, guidance, teaching, messaging, lecturing, email, feedback, review \\ 
        \midrule
        \textbf{Set B} & Time and Space & synchronous, asynchronous, in-person, remote, co-located, hybrid, face-to-face \\ 
        \midrule
        \textbf{Set C} & Agency & AI, Large Language Model (LLM), agent, embodied agent, humanoid agent, bot \\ 
        \bottomrule
    \end{tabularx}
    \label{tab::keywords}
\end{table*}

\begin{figure}[t]
    \centering
    \includegraphics[width=\linewidth]{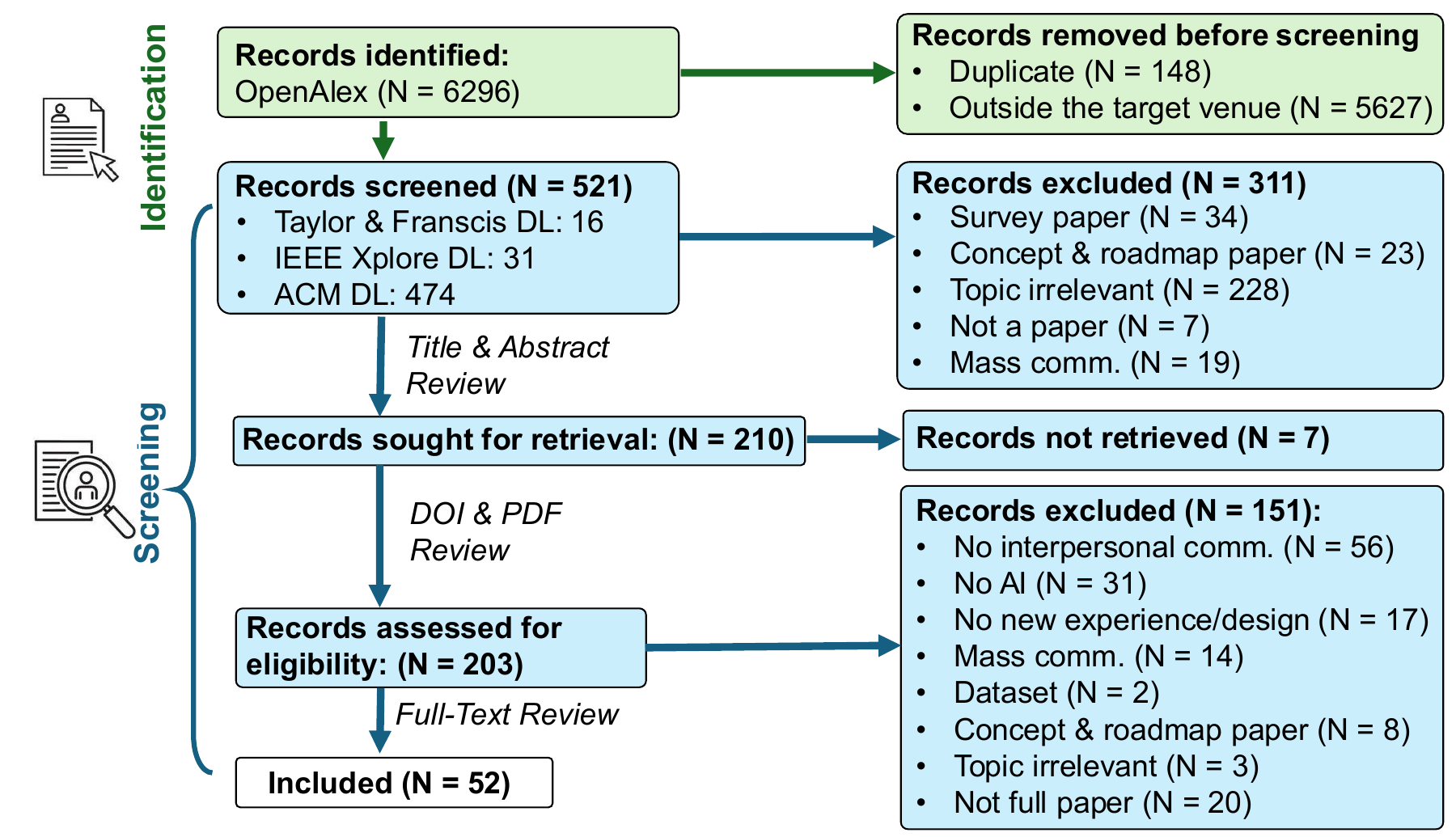}
    \vspace{-.3in}
    \caption{Overview of our review process, including paper counts, following PRISMA.}
    \vspace{-.2in}
    \label{fig::prisma}
\end{figure}

\subsection{Selection Process}\label{sec::methods::selection}

Our selection process follows the established PRISMA framework \cite{page2021prisma, page2022prisma} and consists of two main stages: identification and screening.
Figure~\ref{fig::prisma} illustrates the overall paper selection process.
We first used OpenAlex~\cite{openAlex} to identify candidate records of interest based on the final search query from January 2016 to May 2026, as it provides free API access for paper retrieval without rate limits and aggregates metadata from more than $10$ major academic data sources~\cite{OpenAlexDatasource}.
We then preprocessed the outputs from different formats and consolidated them into a single table. Our identification process yielded $6296$ paper records.
After removing $148$ duplicate records and $5627$ papers not published by the three selected publishers, our identification process yielded $521$ papers.

In the second screening stage, we reviewed the titles and abstracts of the filtered papers and excluded survey papers, studies focusing solely on broad concepts or road maps without concrete research contributions, papers on irrelevant topics, non-article manuscripts, and those centered on mass communication.
This process reduced the dataset to $208$ papers. We then excluded seven papers that were not accessible or searchable.
Through a careful full-text review, we excluded records that met one or more of the following exclusion criteria:

\begin{itemize}[topsep=0pt, itemsep=0pt, leftmargin=9.5pt]

    \item The publication does not include any components of interpersonal communications.

    \item AI is not the primary focus of these publications. For example, some studies, such as MeetScript~\cite{Chen2023MeetScript}, introduce novel meeting visualization systems that rely only on conventional speech-to-text pipelines rather than incorporating AI-driven interaction or analysis features.

    \item The publication focuses on a large-scale study of a specific phenomenon without introducing a novel user experience or tool design~(\eg~\cite{Russo2025}). 

    \item The publication focuses on mass communications (\eg~ through social media).
    
    \item The publication does not report specific research results, for example, when it only introduces a new open-source dataset.

    \item The publication focuses on introducing a new concept rather than reporting specific results. A few full papers published in the full-paper track (\eg~\cite{Rost2026}) may also be considered ineligible under this criterion.

    \item The topic is irrelevant.

\end{itemize}

The authors maintained close communication throughout the screening and eligibility assessment process to discuss and resolve any ambiguities in the included publications. In total, this selection process resulted in a final corpus of $52$ papers.

\subsection{Data Analysis}\label{sec::method::analysis}

We first coded each paper using the two dimensions of the CSCW matrix - \emph{time} and \emph{space}.
We then iteratively recorded several key dimensions that are commonly used to analyze CMC systems.
The final key dimension includes \emph{application use cases}, \emph{communication tasks}, \emph{targeted users}, \emph{targeted size of population}, \emph{devices and displays}.
If the specific paper contains one or multiple studies, we also noted the \emph{study tasks} and \emph{study evaluation methods}.
We finally extracted the \emph{challenges} that are discussed in the papers.
This process enables us to identify four themes of key challenges and research opportunities reported by the authors of the curated papers.
Thematic analysis~\cite{Braun2006}, along with inductive and deductive coding~\cite{Locsin2024}, was used to identify \emph{challenges}, \emph{application use cases}, \emph{study tasks}, and \emph{evaluation} methods.
We conducted our data analysis using Microsoft Excel.

\subsection{Positionally Statement}\label{sec::method::statement}

Our interpretations are shaped by our backgrounds as HCI researchers working across Western academic and industry research contexts.
We include this statement to acknowledge these situated lenses and invite future work that brings additional cultural, methodological, and disciplinary perspectives.
\section{Results}\label{sec::results}

\begin{figure*}
    \centering
    \includegraphics[width=\linewidth]{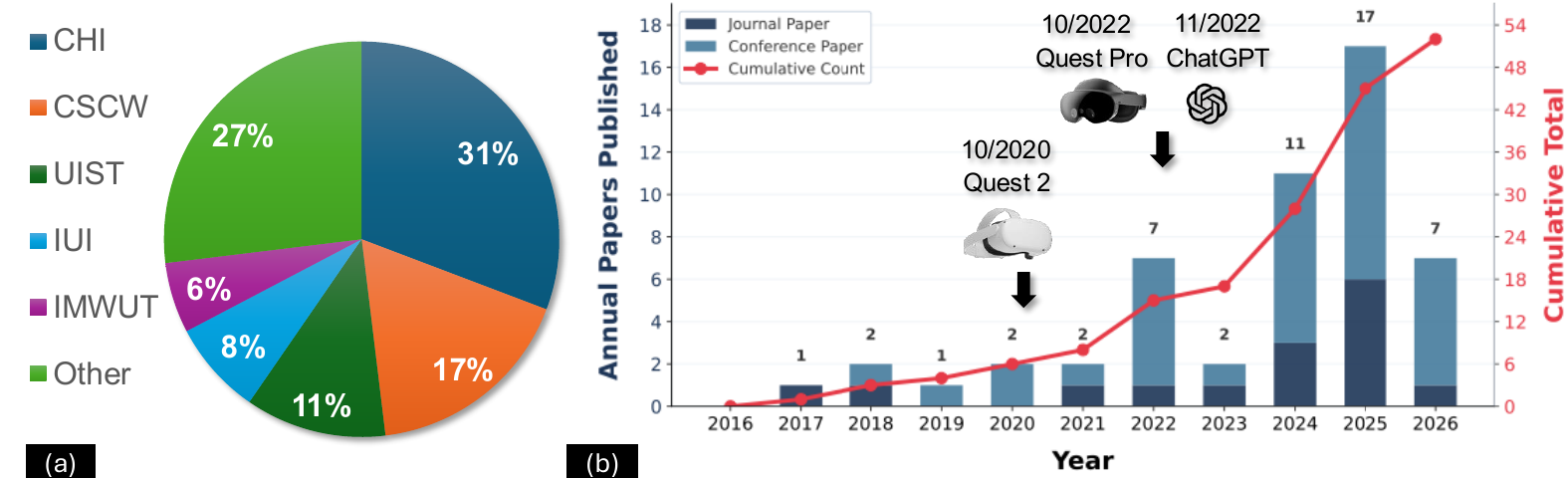}
    \caption{Overview of papers curated through our selection process: (a) distribution across publication venues; (b) publication trends over the last decade. The number of papers published in $2026$ reflects only the first five months of the year.}
    \label{fig::trend}
\end{figure*}

\subsection{Overview of the curated paper corpus}\label{sec::results::characteristics}

Our survey results led to a final corpus of $52$ publications. The complete list of our corpus paper can be referred to Table~\ref{tab::app::literature_review} in Appendix~\ref{sec::app::papers}.

\vspace{4px}\noindent{\bf Trend and venue.}
Figure~\ref{fig::trend}a shows the distribution of publication venues across the selected corpus, with the majority of papers published in CHI (16 papers, $31\%$), CSCW (nine papers, $17\%$), and UIST (six papers, $11\%$).
Figure~\ref{fig::trend}b illustrates the number of curated publications over the past decade.
We also highlighted the emergence of three key enabling technologies that have supported a range of AIMIC research, including Meta Quest 2 (released on October 13, 2020~\cite{Quest2Release}), Meta Quest Pro (released on October 25, 2022~\cite{QuestProRelease}), and ChatGPT (released on November 30, 2022~\cite{ChatGPTRelease}).
Quest 2 was one of the first standalone all-in-one XR headsets~\cite{Quest2Release}, while Quest Pro demonstrated the potential of mixed reality experiences with relatively accessible development~\cite{QuestProRelease}.
These XR headsets have enabled numerous prior studies aimed at designing chat support for in-person conversations (\eg~\cite{Johnson2025}).
On the other hand, ChatGPT has enabled new possibilities for designing agentic experiences powered by pre-trained LLM~\cite{ChatGPTRelease}.

\begin{figure*}
    \centering
    \includegraphics[width=\linewidth]{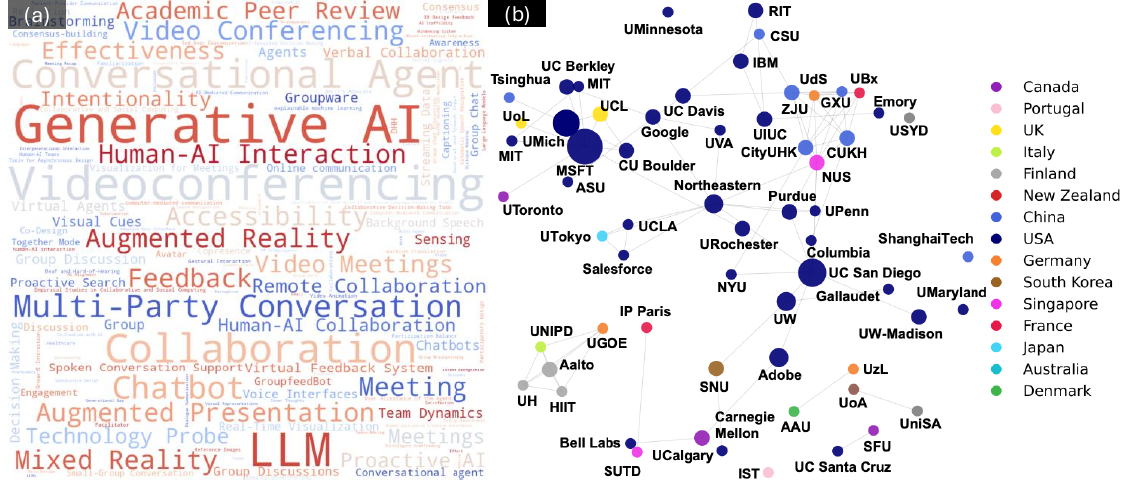}
    \caption{(a) Visualization of author-specified keywords. (b) Overview of the institutional collaboration network in the curated corpus of papers. For multinational industry labs (\eg~Microsoft), we assigned their country based on the location of their headquarters. }
    \label{fig::collaboration}
\end{figure*}

\vspace{4px}\noindent{\bf Contribution types and keywords.}
Among the curated corpus, $17$ papers were published as journal articles, while the remaining $35$ papers appeared in conference proceedings.
Methodologically, our curated corpus reflects a visible historical arc: earlier papers (2017–2021) tend to rely on rule-based agents, classifier pipelines, or WoZ confederates (\eg~IdeaWall~\cite{Shi2017IdeaWall}, CoCo~\cite{Samrose2018a}), while papers from 2023 onward increasingly build on LLMs (especially GPT-3.5/4-class models) both as the underlying reasoning engine and as a generative content source (\eg~MeetMap~\cite{Chen2025MeetMap}, Koala~\cite{Houde2025}).
$49$ papers ($94\%$) contributed both artifacts and empirical studies, while one paper focused solely on an empirical study built on top of an existing tool, and two papers exclusively conducted needs-finding studies.
Functional artifacts contributed by the curated papers include fully functional prototypes as well as low- to medium-fidelity prototypes used as provotypes (\eg~\cite{Johnson2025}), technology probes (\eg~\cite{Park2024CoExplorer}), and prototypes for WoZ studies (\eg~\cite{Rayan2024}).
Figure~\ref{fig::collaboration}a presents a visualization of the 255 keywords provided by the authors. We excluded the keywords from \cite{Rayan2025}, as they did not report any keywords. We also removed generic terms such as ``HCI'', ``AI'', and ``design'' which do not capture the specific focus of the papers.

\vspace{4px}\noindent{\bf Contributors.}
The average number of authors per paper was $5.8$ ($SD = 2.0$), while the average number of affiliated institutions per paper was $2.2$ ($SD = 1.2$).
Figure~\ref{fig::collaboration}b illustrates the collaboration network among affiliated institutions and industry corporations based on the affiliations reported in the selected papers across $15$ countries.
Most of the curated papers were contributed, either fully or partially, by US-based institutions, accounting for $83$ papers ($43\%$).
Among them, Microsoft was identified as the leading institution, contributing to $10$ papers in our curated corpus.

\subsection{Which forms of AIMC are investigated in the current literature? (RQ1)}\label{sec::results::rq1}

\begin{figure*}
    \centering
    \includegraphics[width=0.8\linewidth]{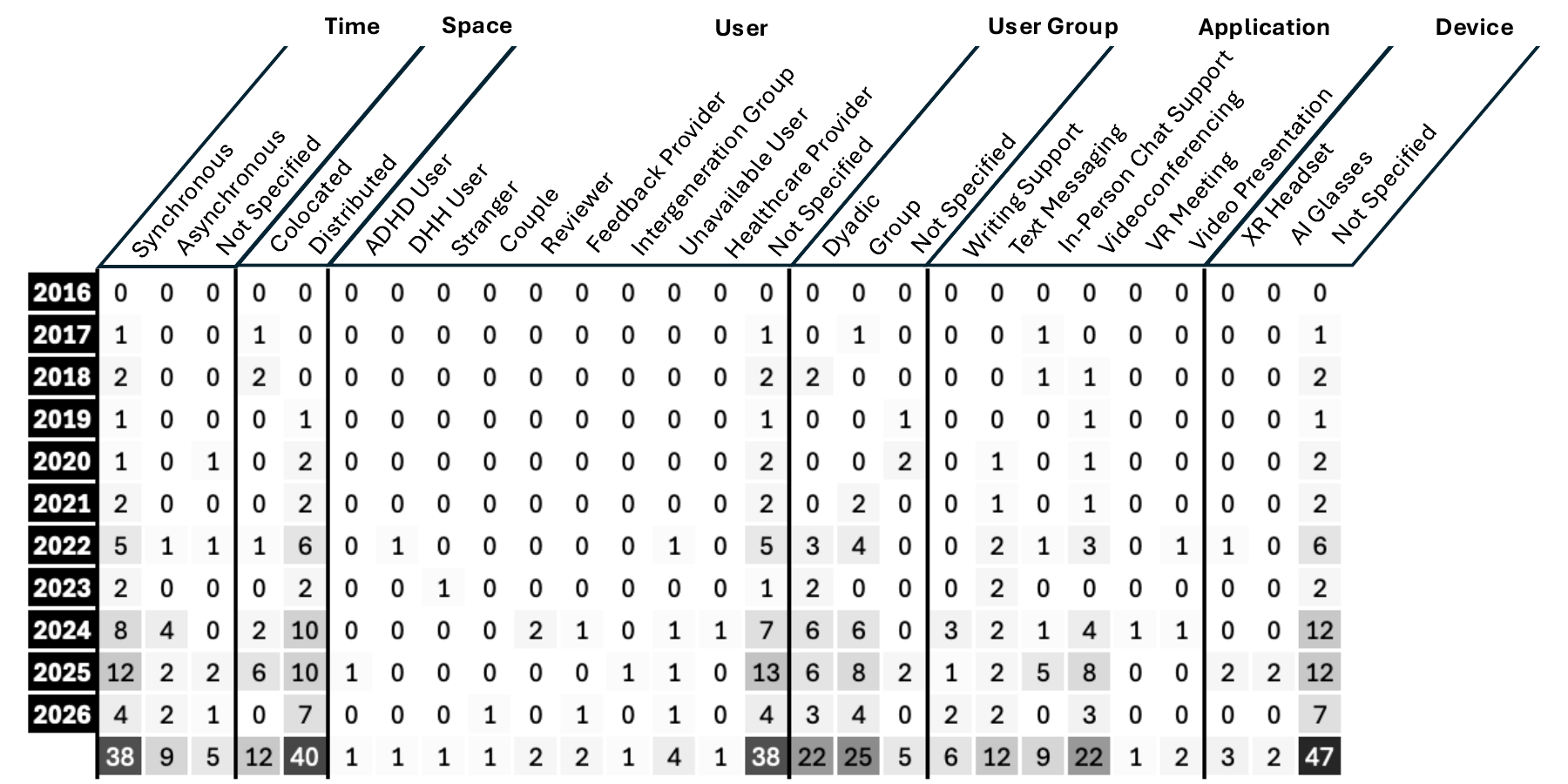}
    \vspace{-0.1in}
    \caption{Overview of the types of AIMIC studied.}
    \vspace{-0.05in}
    \label{fig::type}
\end{figure*}

Figure~\ref{fig::type} presents our analysis of the AIMC types examined in the curated papers. The rest of this section describes our analysis results across six dimensions, including \emph{time}, \emph{space}, \emph{user}, \emph{user group}, \emph{application}, and \emph{device}.

\vspace{4px}\noindent{\bf Time and space.}
We adopted the same \emph{time} and \emph{space} dimensions from the long-standing CSCW matrix to characterize when and where interpersonal communication occurs within the proposed AIMIC experiences.
The majority of papers focus on synchronous ($38$~papers, $73.1\%$) and distributed ($40$~papers, $76.9\%$) interpersonal communication experiences.
In terms of time dimension, most studies have focused on synchronous contexts, such as videoconferencing (\eg~\cite{Aseniero2020MeetCues, Chandrasegaran2019TalkTraces}) and colocated face-to-face chatting (\eg~\cite{Johnson2025, Zhang2025WSCoach, Rayan2024, Rayan2025}).
Several prior works, such as \cite{Wang2026SeeSawBot, Houde2025, Lee2025, Park2024CoExplorer, Kim2020, Liu2025}, did not explicitly specify the time dimension. As these AIMIC systems could potentially support both synchronous and asynchronous interactions, we categorized them as ``not specified.''
We found that these strands of work often focus on understanding the design and role of a facilitating AI agent in group-based text messaging settings.
For example, in group-based text messaging settings, Liu~\etal~\cite{Liu2025} compared three types of proactive agency experiences by designing a traditional reactive agent, a next-speaker prediction model, and a proactive agent with inner thoughts.
Although the study was conducted in a synchronous setting~\cite{Liu2025}, its findings and practical applications can be extended to asynchronous contexts. 
Regarding the spatial dimension, most AIMIC studies conducted in distributed settings have focused on videoconferencing (\eg~\cite{Aseniero2020MeetCues, Chandrasegaran2019TalkTraces}), text messaging (\eg~\cite{Shin2023IntroBot, Wang2026SeeSawBot, Houde2025, Lee2025, Park2024CoExplorer, Kim2020, Liu2025}), and memo-like interpersonal communications such as writing emails (\eg~\cite{Li2025}), providing feedback (\eg~\cite{Chen2024MemoVis, Li2026VizCrit}), and peer reviews (\eg~\cite{Sun2024ReviewFlow, Sun2024MetaWriter}).

\vspace{4px}\noindent{\bf Size of user group.}
Among the papers in our corpus, $22$ ($42.3\%$) focus on dyadic communication, while $25$ ($48.1\%$) focus on group communication. Five papers did not specify the target type of user group.
Most existing work on dyadic communication focuses on in-person chat support (\eg~\cite{Yang2025SocialMind, Zhang2025Understood, Rayan2024, Rayan2025}), videoconferencing (\cite{Valente2022EmpathocAurea, Numan2024SpaceBlender}), and agency for a variety of writing support (\cite{Chen2024MemoVis, Sun2024MetaWriter, Sun2024ReviewFlow, Li2026VizCrit}).
In group settings, five papers focused on group-based text messaging, while seven papers examined the design and affordances of agency in videoconferencing. Only Jonson~\etal~\cite{Johnson2025} explored in-person group conversational experiences.
Five papers did not specify the types of user groups.
For example, while Chen~\etal~\cite{Chen2025AreWeOnTrack} aims to design AI-assisted active and passive goal reflection for videoconferencing experiences, the proposed AIMIC system can be applied to both dyadic and group communication settings.

\vspace{4px}\noindent{\bf Type of users.}
Most of the papers ($38$ papers, $73.1\%$) did not specify the types of users, despite only $14$ papers targeting specific types of users.
Among the papers targeting specific user groups, these included \textbf{general users in particular contexts}; \textbf{users with specialized skills or experience}, such as academic paper reviewers~\cite{Sun2024ReviewFlow, Sun2024MetaWriter}, feedback providers for visual~\cite{Li2026VizCrit} and 3D design~\cite{Chen2024MemoVis}, strangers collaborating on shared tasks~\cite{Shin2023IntroBot}, healthcare providers~\cite{Bedmutha2024}, relational groups like couples~\cite{Jiang2026ScaffoldedVulnerability}, and intergeneration groups~\cite{Kim2025Generations}; and \textbf{minority populations}, including users with \textbf{A}ttention-\textbf{D}eficit/\textbf{H}yperactivity \textbf{D}isorder (ADHD)~\cite{Zhang2025Understood} and \textbf{D}eaf and \textbf{H}ard of \textbf{H}earing (DHH) users~\cite{McDonnell2021}.
Among the papers targeting general user populations, several studies highlighted the affordances and potential value of these systems for specific user groups. 
For example, although SocialMind~\cite{Yang2025SocialMind} introduced a proactive LLM-based conversational assistant for dyadic interactions, they also identified its potential applications in supporting individuals with \textbf{S}ocial \textbf{A}nxiety \textbf{D}isorder (SAD) and \textbf{A}utism \textbf{S}pectrum \textbf{D}isorder (ASD).

\vspace{+4px}\noindent{\bf Device.}
The majority of the surveyed papers did not explicitly specify the devices required to support the proposed AIMIC experience. Notably, our analysis did not treat standard computing devices, such as desktop computers and smartphones, as dedicated devices.
Among our curated corpus papers, five papers identified the needs of introducing additional hardware to realize AI-mediated synchronous and colocated interpersonal conversation.
Three papers leverage MR headsets to render virtual supporting information, including text-based cues~\cite{Zhang2025Understood}, embodied agents~\cite{Johnson2025}, and visualizations of the cognitive states of communication partners.
Two papers leveraged lightweight AI glasses. SocialMind~\cite{Yang2025SocialMind} employs RayNEO X2~\cite{RayNeoGlassesRelease2023, RayNeoGlasses2024}, a programmable pair of AI glasses with a built-in display (Figure \ref{fig::headset}a - b), while WSCouch~\cite{Zhang2025WSCoach} uses Huawei Eyewear~\cite{HuaweiEyeWear}, a lightweight AI glasses without a display, to help users reduce speech disfluencies through a novel auditory intervention framework.

\begin{table*}[htbp]
\centering
\caption{Taxonomy of application domains and sub-categories. Some surveyed papers may address multiple tasks within a single application context.}
\label{tab:application_taxonomy}
\vspace{+4px}
\begin{tabular}{p{0.48\textwidth}|p{0.48\textwidth}}

\toprule
\textbf{Writing Support} & \textbf{Text Messaging} \\
$\rightarrow$ Assist creating feedback for 2D visual design~\cite{Li2026VizCrit} & $\rightarrow$ Facilitate conversation~\cite{Chiang2024, Liu2025, Houde2025} \\
$\rightarrow$ Assist creating 3D design feedback~\cite{Chen2024MemoVis} & $\rightarrow$ Cross-private and shared channel support~\cite{Wang2026SeeSawBot} \\
$\rightarrow$ Assist email writing~\cite{Li2025, Yao2026PersonaMail} & $\rightarrow$ Support self-disclosure interaction~\cite{Jiang2026ScaffoldedVulnerability} \\   
$\rightarrow$ Assist writing peer review feedback~\cite{Sun2024MetaWriter, Sun2024ReviewFlow} & $\rightarrow$ Support cognitive and social awareness~\cite{deJong2024} \\   

\midrule
\textbf{In-Person Chat Support} & \textbf{Videoconferencing} \\
$\rightarrow$ Suggestive information support~\cite{Yang2025SocialMind, Zhang2025Understood} & $\rightarrow$ Feedback \& Reflection~\cite{Chen2025AreWeOnTrack, Scott2025WhatDoesSuccessLookLike, Asthana2025} \\
$\rightarrow$ Inspiration for conversation~\cite{Andolina2018} & $\rightarrow$ Partial or no participation~\cite{Leong2024, Bai2026PartialParticipation}  \\
$\rightarrow$ Reduction of unwanted words~\cite{Zhang2025WSCoach} & $\rightarrow$ Enable shared task space~\cite{Rajaram2024BlendScape}  \\
$\rightarrow$ Facilitate group conversation~\cite{Johnson2025} & $\rightarrow$ Meeting with avatar(s) and embodied agent(s)~\cite{Leong2024, Panda2022}  \\
$\rightarrow$ Patient-doctor communication~\cite{Bedmutha2024, Samiee2025}  & $\rightarrow$ Cognitive augmentation~\cite{Suzawa2025}  \\
$\rightarrow$ Support for disabled users~\cite{McDonnell2021} & $\rightarrow$ Support for disabled users~\cite{Seita2022}  \\
$\rightarrow$ Cognitive augmentation~\cite{Valente2022EmpathicAuRea} & $\rightarrow$ Promote more inclusive and smoother meetings~\cite{Houtti2025ObserveAskIntervene, Rayan2024, Rayan2025, Johnson2026} \\

\midrule
\textbf{VR Meeting} & \textbf{Video Presentation} \\
$\rightarrow$ Enable shared task space~\cite{Numan2024SpaceBlender} & $\rightarrow$ Augment video presentation~\cite{Liao2022RealityTalk} \\

\bottomrule
\end{tabular}
\label{tab::tasks}
\end{table*}

\vspace{4px}\noindent{\bf Application and Task.}
Despite the wide variety of interpersonal communications, we found that prior works primarily focus on six types of application contexts (Figure~\ref{fig::type}). 
Each paper often focuses on specific tasks and interaction challenges contextualized on the specific application (Table~\ref{tab::tasks}). 
Prior literature on AI-assisted writing support tools has primarily focused on feedback-oriented applications, including peer review~\cite{Sun2024MetaWriter, Sun2024ReviewFlow}, feedback for visual~\cite{Li2026VizCrit} and 3D design~\cite{Chen2024MemoVis}, and email writing~\cite{Li2025, Yao2026PersonaMail}.
These applications often aim to develop novel human-AI collaborative workflows that facilitate the creation of asynchronous communication messages more efficiently and with higher quality.
For example, MemoVis~\cite{Chen2024MemoVis} demonstrates how a range of vision-language foundation models can be integrated to help 3D design feedback providers generate higher-quality reference images that can be incorporated into design feedback.
Rather than focusing on aesthetics, Chen~\etal~considered high-quality reference images as those that effectively visualize the intent of textual feedback without introducing or altering design elements not explicitly mentioned in the feedback~\cite{Chen2024MemoVis}.
Regarding the application of text messaging, our analysis identified five tasks that prior surveyed paper are focusing on. 
While most prior research focuses on understanding the design, feasibility, and affordances of AI agents in text messaging contexts, we also identified studies that target more specific tasks for particular user groups.
For example, Jiang~\etal~\cite{deJong2024} explored the use of an AI chatbot to support self-disclosure and need-based supportive communication between couples through a dual-layer vulnerability scaffolding framework.
SeaSawBot~\cite{Wang2026SeeSawBot} explores how AI agents can support communication across private and public channels in IM applications such as Slack, with the goal of enhancing team dynamics and collaboration.
Applications of in-person chat support focus on colocated, synchronous conversations augmented by AI through various forms of information intervention, such as \insitu~visual support delivered via MR headset~\cite{Yang2025SocialMind, Valente2022EmpathocAurea} or AI-enabled glasses with displays~\cite{Yang2025SocialMind}, as well as auditory cues provided through wearable AI devices~\cite{Zhang2025WSCoach}.
A variety of tasks have been explored in the context of video conferencing. However, most surveyed papers focus on designing real-time feedback mechanisms and AI-assisted reflection tools during different phases of the video meeting~\cite{Chen2025AreWeOnTrack, Scott2025WhatDoesSuccessLookLike, Asthana2025}, as well as exploring new approaches to encourage participation and foster more inclusive meetings for collaborative group decision-making tasks~\cite{Chen2023MeetScript, Houtti2025ObserveAskIntervene}.
A few studies have explored less commonly examined applications, such as VR-based meetings (\eg~\cite{Numan2024SpaceBlender}) and video presentation contexts (\eg~\cite{Liao2022RealityTalk}).

\begin{tcolorbox}[enhanced, colback=darkblue!10, colframe=white, boxrule=0pt, arc=3pt, fuzzy shadow={3.5pt}{-3.5pt}{0pt}{0.4pt}{black!20}]

\textcolor{darkblue}{\textbf{Key Takeaway:}}
Prior AIMIC research primarily focuses on synchronous, distributed settings such as videoconferencing and text messaging, while most papers leave user type, user group, and device unspecified. 
Group communication is studied nearly as often as dyadic communication, yet few systems explicitly target minority populations such as DHH or ADHD users. 
Standard computing devices remain the default; dedicated hardware like XR headsets and AI glasses appears only in a small, recently emerging subset of colocated designs. 
\end{tcolorbox}

\subsection{How has AI been integrated into different forms of CMC, and how can these approaches be systematically organized? (RQ2)}\label{sec::results::rq2}

This section presents our analysis results across four dimensions: \emph{AI embodiment}, \emph{AI shareability}, \emph{initiator and message sender}, and \emph{AI techniques being applied}.

\begin{figure*}[t]
    \centering
    \includegraphics[width=\linewidth]{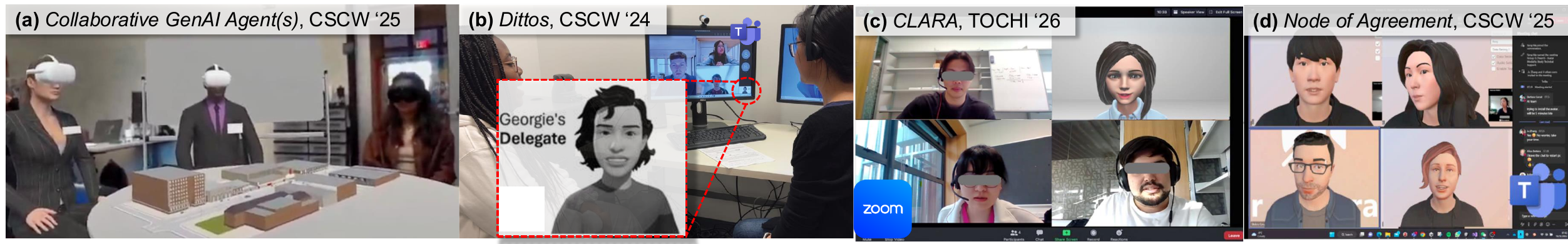}
    \vspace{-.3in}
    \caption{Selected examples of the AIMIC experiences from the curated literature. Figures are taken from \cite{Johnson2025}, \cite{Leong2024}, \cite{Gunasekaran2026CLARA}, and \cite{Ma2025}, respectively, from left to right.}
    \label{fig::embodiedagentexample}
\end{figure*}

\vspace{4px}\noindent{\bf AI Embodiment.}
We use \emph{AI embodiment} to refer to how AI was designed in a specific AIMIC experience.
Overall, $16$ papers used AI to provide various forms of information visualization support. 
$13$ papers designed AI-driven chatbots to facilitate interpersonal communication.
While most prior research (\eg~\cite{Shin2022, Kim2020}) has designed and prototyped chatbots within customized web applications, some recent works, such as Koala~\cite{Houde2025}, have further integrated LLM-powered assistive chatbots into existing commercial instant messaging platforms, such as Slack.
Six papers focused on AI-powered embodied agents. 
Figure~\ref{fig::embodiedagentexample} illustrates example AIMIC experiences from the curated literature in in-person (Figure~\ref{fig::embodiedagentexample}a) and distributed settings (Figures~\ref{fig::embodiedagentexample}b - d).
For example, Johnson~\etal~\cite{Johnson2025} explores the role of embodied agents in facilitating in-person group meetings (Figure~\ref{fig::embodiedagentexample}a). 
At the same time, Ditto~\cite{Leong2024} investigates the design of a delegate agent - a humanoid embodied representation of an unavailable meeting participant - in remote videoconferencing settings (Figure~\ref{fig::embodiedagentexample}b).
While most papers advocate for embodied agents driven by large AI models, few (\eg~\cite{Ma2025, Panda2022}) explore agents operated by real human users. 
For example, Ma~\etal~\cite{Ma2025} examine video meeting outcomes when an animated avatar is driven by a live webcam feed (Figure~\ref{fig::embodiedagentexample}d).
$17$ papers introduced dedicated AI-assisted companion tools.

\vspace{4px}\noindent{\bf Shareability.}
Shareability refers to the extent to which the AIMIC experience can be shared among communication participants.
We categorize existing research along this dimension into \emph{private}, \emph{partially shared}, and \emph{shared}.
A \emph{private} AIMIC experience refers to those altered and/or mediated by AI that are only visible to the supported communication participant(s).
Our analysis identified $23$ papers that explored the design of private AIMIC systems.
The key application scenarios include tools that support videoconferencing participants (\eg~\cite{Aseniero2020MeetCues, Bai2026PartialParticipation}) and systems that leverage various display technologies to facilitate colocated synchronous communication (\eg~\cite{Andolina2018, Yang2025SocialMind, Zhang2025Understood, Bedmutha2024}).
In contrast, a \emph{shared} AIMIC experience refers to AI-altered and/or AI-mediated interactions that are visible and accessible to all communication participants.
$20$ papers that explored the design of shared AIMIC experience.
Examples include shared chatbots in text messaging applications~\cite{Kim2020, Shin2022}, shared embodied agents in collaborative group tasks~\cite{Johnson2025, Panda2022, Leong2024, Johnson2026}, and shared task space in remote meeting experiences, such as backgrounds~\cite{Rajaram2024BlendScape} in videoconferencing experiences, as well as shared 3D space in immersive VR meetings~\cite{Numan2024SpaceBlender}.
The AIMIC systems explored in five papers were considered to support both private and shared communication experiences.
For example, FacilitatorBot~\cite{Do2023} was designed to detect under-contributing members in group text chat settings and send private supervisory messages, while also broadcasting task-related information and sending reminders about meeting times.
Another example is SeeSawBot~\cite{Wang2026SeeSawBot}, which explores the use of both private and public channels to support group chat dynamics.

\vspace{4px}\noindent{\bf Initiator and Message Sender.}
We adopted the four design patterns summarized by Lee~\etal~\cite{Lee2025} and categorized them through the lens of initiator (\ie~who initiates the request for AI assistance) and message sender (\ie~who sends the communication messages (Figure~\ref{fig:teaser}). The design paradigm surrounding the initiation of AI assistance can also be understood through the long-standing theory of mixed-initiative interaction design~\cite{Novick1997, Horvitz1999}.
We consider that both the initiator and the sender of communication messages can be either \emph{human} or \emph{AI}. In some papers, both humans and AI may simultaneously serve as initiators and message senders.
Overall, we identified $11$ papers in which AIMIC is initiated by humans, while the communication messages are ultimately relayed, integrated, and delivered by humans to their communication partners. Nearly all of these papers position AI as a companion tool that users can leverage to reformulate communication messages before sharing them with others. For example, MetaWriter~\cite{Sun2024MetaWriter} and ReviewFlow~\cite{Sun2024ReviewFlow} proposed new AI-mediated tools to assist peer reviewers in writing review comments.
Four papers adopted a design in which AIMIC is initiated by humans, while AI reformulates the messages before forwarding them to communication partners.
For instance, BlendSpace~\cite{Rajaram2024BlendScape} designed an AI tool capable of automatically creating a ``blended'' shared background for videoconferencing that is jointly shared among conversation partners.
$33$ papers explore the use of AI as the initiator of AIMIC experience.
Among these AI-initiated designs, 23 relied on humans to deliver the communication messages, while 10 relied on AI to send the messages directly.
Deciding how and when to trigger the AI support by leveraging heterogeneous complex context has long been considered a challenging problem in designing a broader proactive system~\cite{Deng2025, Fischer2012}.
Our analysis further identified three strategies used in prior research to determine when to initiate AI support: continuously streaming AI-inferred assistance (\eg~\cite{Andolina2018, Rajaram2024BlendScape}), applying predefined rules based on factors such as critical timing (\eg~\cite{Kim2020}), participant contributions (\eg~\cite{Kim2020}), gestural behaviors (\eg~\cite{Cao2024Elastica, Liao2022RealityTalk}), and cognitive states (\eg~\cite{Gunasekaran2026CLARA}), as well as leveraging the reasoning capabilities of task-specific AI models or large foundation models~\cite{Houde2025, Liu2025}.

\vspace{4px}\noindent{\bf AI techniques employed.}
By analyzing the implementations described in the surveyed papers, we identified five categories of AI techniques, including traditional NLP algorithms, AI techniques to understand non-verbal cues, language models, vision-related AI techniques, and AI techniques for 3D.
Notably, our analysis did not include standard transcription and speaker diarization algorithms, as these techniques are almost universally adopted across nearly all AIMIC research.
Some curated papers may be labeled as employing multiple AI techniques while prototyping the specific interactive experiences.
$11$ papers adopted traditional NLP techniques, such as key entity extraction (\eg~\cite{Shi2017IdeaWall, Andolina2018}), lexical and morpheme analysis (\eg~\cite{Kim2020}), and topic modeling (\eg~\cite{Chandrasegaran2019TalkTraces, Shin2023IntroBot}), to support a variety of AIMIC experiences.
All of these papers were published in or before 2023.
$22$ papers used language models, such as BERT and GPT-based LLMs, while prototyping their AIMIC experiences, all of which were published in 2024 or later.
Five papers employed specialized AI techniques to understand various non-verbal cues, such as emotions and gestural behaviors, and a variety of social signals.
Four papers employed vision-related AI techniques, such as inpainting and ControlNet, to support applications including the creation of blended shared videoconferencing backgrounds~\cite{Rajaram2024BlendScape} and the generation of reference images for design feedback~\cite{Chen2024MemoVis}.
Four papers employed AI techniques for 3D applications, such as 3D Gaussian Splatting~\cite{Hu2025} and the integrated use of semantic segmentation, depth estimation, and backprojection to blend multiple 3D scenes for immersive VR meetings~\cite{Numan2024SpaceBlender}.

\begin{tcolorbox}[enhanced, colback=darkblue!10, colframe=white, boxrule=0pt, arc=3pt, fuzzy shadow={3.5pt}{-3.5pt}{0pt}{0.4pt}{black!20}]

\textcolor{darkblue}{\textbf{Key Takeaway:}}
Across AI embodiment, shareability, initiator/sender roles, and underlying technique, the field is shifting from rule-based NLP pipelines used before 2023 toward LLM-powered agents capable of greater autonomy. 
Most systems still have AI initiated by a human, but a growing share let AI both decide when to intervene and deliver messages to communication partners directly, without a human relay. 
Deciding when to trigger AI assistance remains an unresolved design problem. 
Prior research has explored the method of continuous streaming, predefined rules, or model-based reasoning over context. 
This variation in initiation strategy, more than the specific AI technique used, appears to be the primary driver of differences in reported user experience.

\end{tcolorbox}

\subsection{What are the outcomes and benefits for the AIMIC experiences explored in current literature? (RQ3)}\label{sec::results::rq3}

Our scoping review identified seven categories of key benefits that the curated publications aim to achieve when exploring existing AIMIC experiences and/or designing new AIMIC systems. It is worth noting that some publications are associated with multiple benefit categories in relation to the AIMIC systems they investigate. 
The top two benefits that our analysis identified are intervene and facilitate ($18$ papers, $34.6\%$) and providing \insitu~information support ($17$ papers, $32.7\%$).

\vspace{4px}\noindent{\bf Providing \insitu~information support}.
$17$ papers explored the design of AIMIC systems to provide in-situ support for a range of interpersonal communication experiences, including eight papers focused on co-located synchronous conversations and seven examining various forms of distributed communication.
While most papers do not explicitly specify their target users, two studies focus on DHH populations~\cite{McDonnell2021, Seita2022}.
Publications in this category often explore the use of AI to interpret complex conversational dynamics and provide in-situ information support to facilitate and enhance communication experiences.
$13$ papers explored the use of AI to augment and visualize communication signals, such as subtle back-channel cues (\eg~\cite{Valente2022EmpathicAuRea}), while four papers aimed to provide additional details grounded in prior conversational context (\eg~\cite{Andolina2018}).
We identified five types of information explored across the curated publications, including private and shared collaborative visualization for in-depth understanding of communication messages (\eg~meeting topics~\cite{Andolina2018, Aseniero2020MeetCues, Chandrasegaran2019TalkTraces}, translated messages in cross-lingual communication settings~\cite{Robertson2022}, and captions for DHH users~\cite{McDonnell2021, Seita2022}), information support to inspiring for new communication messages (\eg~\cite{Andolina2018, Shi2017IdeaWall}), additional overlays that are adaptively added to the augmented video presentation~\cite{Liao2022RealityTalk, Cao2024Elastica}, and visualized information that augments cognitive states of communication partners~\cite{Valente2022EmpathicAuRea}.
One instantiation of the \insitu~information support is the design of in-meeting awareness and sense-making tools that attempt to make the flow of videoconferencing-based discussion easier to understand. 
For example, MeetMap~\cite{Chen2025MeetMap} designed two LLM-assistance levels, \emph{human-map} (AI drafts summary nodes, humans arrange them) and \emph{AI-map} (AI drafts the whole map, humans edit), and unveiled the benefits that both outperformed existing Zoom-plus-AI-summary baseline on comprehension, with AI-map preferred for low effort and human-map preferred when participants wanted sense-making agency.
We have also observed how recent advances in large foundation AI models have reshaped research opportunities for providing \insitu~ informational support. For example, earlier work such as \cite{Chandrasegaran2019TalkTraces, Shi2017IdeaWall, Andolina2018} relied on topics and key entities extracted through traditional NLP techniques to help users better understand meetings, whereas more recent research such as \cite{Rayan2025} has explored the use of LLMs to interpret meeting dynamics more deeply.

\vspace{4px}\noindent{\bf Establish a shared context.}
Six papers explored the use of AI to augment the shared context.
Papers in this category often aim to establish shared informational grounding to augment remote interpersonal communication experiences.
While three papers~\cite{Numan2024SpaceBlender, Rajaram2024BlendScape, Hu2025} focus on distributed dyadic conversations, we believe their innovative ideas and interaction designs can be extended to broader, more complex group scenarios.
Three papers explored the use of embodied agents as the shared representation for the videoconferencing participants~\cite{Panda2022, Leong2024, Ma2025}.
Three papers explored the use of large vision-language foundation models to establish a shared task space~\cite{Numan2024SpaceBlender, Rajaram2024BlendScape, Hu2025}.

\vspace{4px}\noindent{\bf Intervene and facilitate.}
$17$ explored the use of AI to intervene and facilitate a variety of interpersonal communications.
Among the papers in this category, $13$ explored the use of chatbots in text-messaging settings, three investigated embodied agents, and one examined an embodied agent in an in-person MR-mediated group conversation context.
The majority of papers explored group communication scenarios, with only two focusing on dyadic interactions. Most papers did not specify target populations, with only Jiang~\etal~\cite{Jiang2026ScaffoldedVulnerability} examining conversations between couples.
%
% What can the LLM agent do in the group discussion?
Our analysis identified seven benefits explored in prior publications, including the use of LLM agents to encourage participation from less active group members~\cite{Kim2020, Houtti2025ObserveAskIntervene, Johnson2026}, leveraging AI to summarize conversational context when needed~\cite{Kim2020}, promoting group consensus building~\cite{Shin2022, Chiang2024}, designing AI agents as ``devil’s advocates'' to stimulate debate, test opposing arguments, and encourage the exploration of diverse perspectives~\cite{Chiang2024}, enhancing communication fluency~\cite{Zhang2025WSCoach}, manage group dynamics~\cite{Wang2026SeeSawBot, deJong2024} and supporting self-disclosure and needs-based support~\cite{Jiang2026ScaffoldedVulnerability}.

\vspace{4px}\noindent{\bf Participation in communication when unavailable.}
Two papers explored how AI can facilitate meeting participation when users are partially or fully unavailable.
While Ditto~\cite{Leong2024} proposes the use of an AI-driven embodied delegate agent to participate in videoconferencing, ProxyMe~\cite{Bai2026PartialParticipation} focuses on how an AI agent can support knowledge workers in partially participating in meetings by providing LLM-generated topic summaries and semi-automatically responding to questions.

\vspace{4px}\noindent{\bf Augment interaction workflows.}
Eight papers use AI to enable a range of interaction workflows that support more efficient and effective interpersonal communication experiences.
Papers in this category often position AI as a companion tool within communication-enabling applications.
Two papers used AI to support planning and reflection in videoconferencing applications. 
Two papers used AI to assist peer reviewers in writing higher-quality reviews~\cite{Sun2024ReviewFlow, Sun2024MetaWriter}. 
Two papers used AI to help knowledge workers compose emails~\cite{Li2025, Yao2026PersonaMail}. 
Finally, two papers used AI as a companion tool to augment the workflow of creating design feedback~\cite{Chen2024MemoVis, Li2026VizCrit}.

\vspace{4px}\noindent{\bf Reflection and feedback.}
Four papers explored the use of AI to nudge communication participants to engage with intentional reflection and feedback.
All four papers focus on videoconferencing applications, with two using AI to facilitate post-meeting feedback~\cite{Samrose2018a, Samrose2021}, two exploring AI support for in-meeting reflection~\cite{Chen2025AreWeOnTrack}, and one examining prospective reflection to help participants clarify why a meeting is needed and what may occur~\cite{Scott2025WhatDoesSuccessLookLike}.
While \emph{reflection} is sometimes considered a form of \emph{intrapersonal communication} - which refers to the active process of communicating with oneself through internal dialogues, thoughts and reflections~\cite{Bainbridge2025, Cunningham1992} - our analysis treats it as a process that enhances the delivery and comprehension of communication messages.

\begin{tcolorbox}[enhanced, colback=darkblue!10, colframe=white, boxrule=0pt, arc=3pt, fuzzy shadow={3.5pt}{-3.5pt}{0pt}{0.4pt}{black!20}]

\textcolor{darkblue}{\textbf{Key Takeaway:}}
Providing \insitu~information support and actively intervening to facilitate communication are by far the most common goals, together motivating over two-thirds of the systems in our corpus. 
Benefit categories are not mutually exclusive: many systems pursue several simultaneously, such as pairing in-situ support with workflow augmentation. 
Fewer systems target enabling participation when a user is unavailable or prompting reflection, despite both being promising directions for asynchronous and hybrid work.
Notably, nearly every facilitation- or reflection-oriented system also reports a cost in cognitive load or perceived agency.

\end{tcolorbox}

\subsection{What are the key challenges and opportunities identified? (RQ4)}\label{sec::results::challenges}

\begin{figure*}[t]
    \centering
    \includegraphics[width=\linewidth]{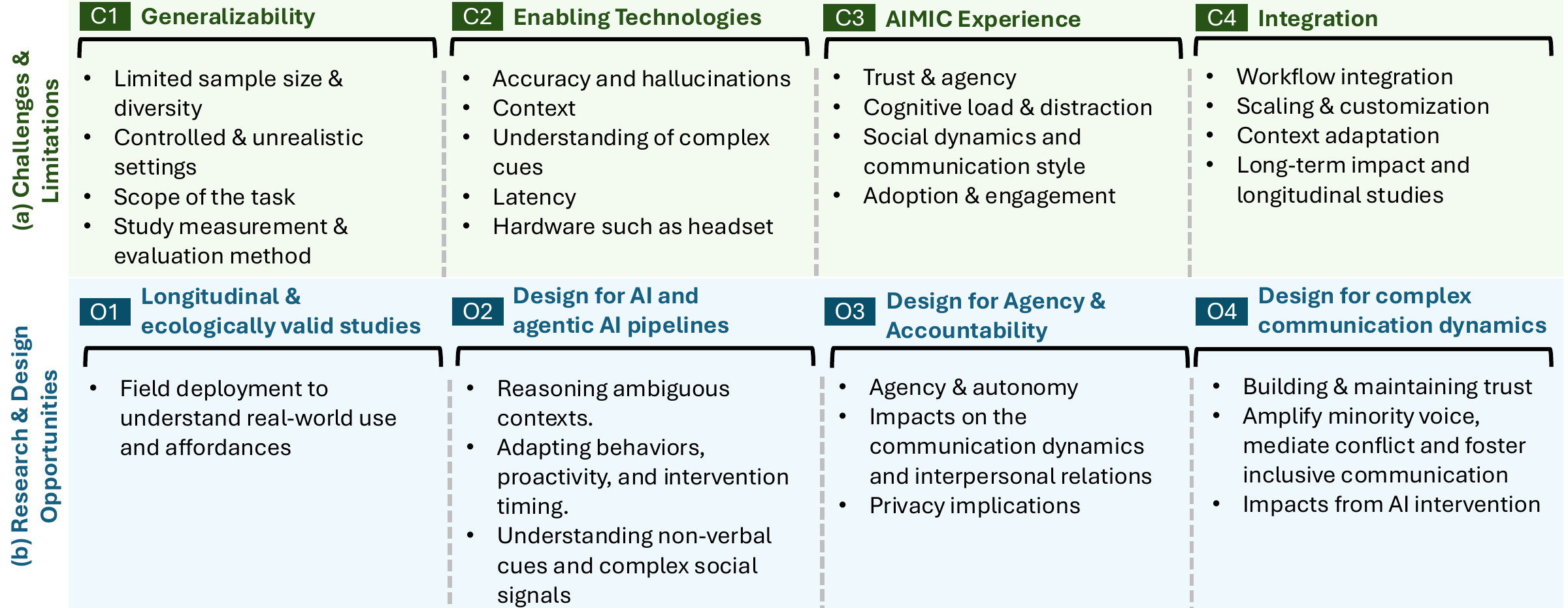}
    \vspace{-.3in}
    \caption{Overview of the main themes and subthemes of (a) challenges and limitations, and (b) research and design opportunities.}
    \vspace{-.1in}
    \label{fig::challenges}
\end{figure*}

Figure~\ref{fig::challenges} presents an overview of the primary themes and subthemes of emerging challenges and limitations (Figure~\ref{fig::challenges}a), as well as the research and design opportunities (Figure~\ref{fig::challenges}b) identified across our paper corpus.

\vspace{4px}\noindent{\bf Challenges and limitations.}
Overall, the identified challenges and limitations span four primary themes (Figure~\ref{fig::challenges}a). We use \colorbox{green!15}{C\#} to index each key challenge and limitation.

\vspace{4px}\noindent {\bf \colorbox{green!15}{C1}~Methodological and generalizability.}
Our analysis identified key limitations related to small sample size and participant diversity, as well as the controlled and unrealistic settings.
Most of the reviewed papers relied on small and homogeneous participant pools (\eg~\cite{Li2025}), such as university students or employees from a single company (\eg~\cite{Chen2025AreWeOnTrack}).
Papers employing technology probe approaches, such as \cite{Wang2026SeeSawBot, Park2024CoExplorer, Chen2025AreWeOnTrack}, also highlighted limitations stemming from the complexities of real-world text messaging and videoconferencing communication settings.
Our analysis also identified a third limitation related to the narrow scope of supported tasks and the limited range of AI mediation approaches.
Finally, we found that difficulties in quantitatively assessing the true impacts of AI mediation emerged as another key limitation highlighted by the authors.
For example, Valente~\etal~\cite{Valente2022EmpathicAuRea} pointed out concerns regarding the reliability of self-reported metrics, as well as challenges associated with the emotion recognition pipeline.

\begin{figure*}[t]
    \centering
    \includegraphics[width=\linewidth]{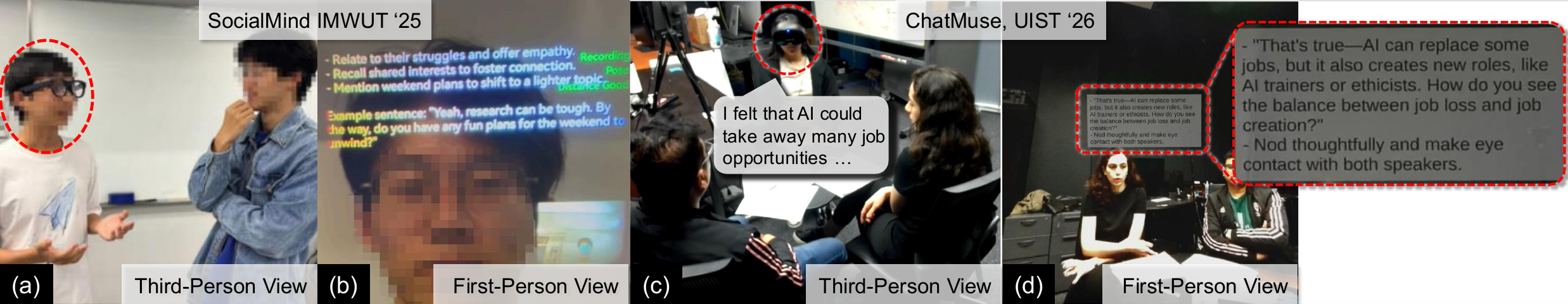}
    \vspace{-.3in}
    \caption{Example prior studies focusing on in-person interpersonal conversation with (a - b) AI glasses and (c - d) MR headset. Figures shown in (a) and (b) are taken from \cite{Yang2025SocialMind} and \cite{Zhou2026ChatMuseEA, Zhou2026ChatMuse}, respectively. Individuals wearing the head-mounted devices are highlighted with red circles.}
    \vspace{-.1in}
    \label{fig::headset}
\end{figure*}

\vspace{4px}\noindent {\bf \colorbox{green!15}{C2} Limitation of existing AI techniques and enabling technologies.}
The integrated AIMIC experience has frequently been reported to produce inaccurate or misleading outputs, which may frustrate communication participants and undermine trust in the system.
While many recent studies have explored the feasibility of integrating additional non-verbal cues into AIMIC systems~\cite{Yang2025SocialMind}, existing AI techniques are still often challenged in accurately understanding rich multimodal information, which is widely regarded as a critical component of interpersonal communication.
Latency has also been frequently cited as another key limitation~\cite{Bai2026PartialParticipation, Cao2024Elastica, Yang2025SocialMind, Johnson2025, Leong2024}.
This limitation becomes particularly pronounced in group interaction settings (\eg~\cite{Johnson2025, Leong2024}) as well as in AIMIC experiences that rely on vision-based models (\eg~\cite{Numan2024SpaceBlender, Rajaram2024BlendScape}).
In the context of in-person AIMIC settings supported by XR headsets, a few studies, \eg~\cite{Yang2025SocialMind, Zhang2025Understood}, have identified limitations associated with the headset itself, which can obstruct critical non-verbal communication channels and introduce additional hardware constraints (\eg~comfort, weight, and battery life, which may affect long-term use).
Figure~\ref{fig::headset} illustrate participants receiving communication support while wearing lightweight AI glasses (\eg~SocialMind~\cite{Yang2025SocialMind}) and a Quest Pro headset (\eg~\cite{Zhou2026ChatMuse}), respectively.

\vspace{4px}\noindent {\bf \colorbox{green!15}{C3}~AIMIC experience.}
Our analysis has noted that users often express skepticism, mistrust, or an unwillingness to cede control to AI, particularly when its workings are opaque or its output is unreliable.
There is also a risk of overreliance, where users may engage with the AIMIC system in ways that differ from the original design intentions. 
For example, Scott~\etal~\cite{Scott2025WhatDoesSuccessLookLike} reported that some participants preferred receiving immediate solutions, even though the designed AI-mediated experience was intended to encourage reflection.
These lines of inquiry highlight the persistent challenge of seeking ``calibrated trust'' based on the interplay between human trust and AI competence, which has been recognized as a long-standing hurdle in designing aligned human-centered AI systems~\cite{Shneiderman2020, Lee2004}.
A second challenge is related to the increased cognitive load, distractions, and unintentional disruptions to the natural communication flow.
Scott~\etal~\cite{Scott2025WhatDoesSuccessLookLike} note that reflection can entail significant time costs. Park~\etal~\cite{Park2024CoExplorer} raise concerns about potential distractions caused by frequent prompts. Bai~\etal~\cite{Bai2026PartialParticipation} highlight a ``productivity paradox,'' in which AI mediation may paradoxically increase cognitive load. Similarly, Ryan~\etal~\cite{Rayan2025} acknowledge that consuming and interpreting cues designed to support co-located, in-person communication can instead introduce additional cognitive load and become distracting when misaligned with user needs.
The third challenge concerns the impact on social dynamics, communication styles, and sometimes even fundamental communication behaviors.
For example, Johnson~\etal~\cite{Johnson2025} highlight how humanoid GenAI agents may disrupt team dynamics by introducing concerns about surveillance and potentially weakening interpersonal relationships.
The final challenge in this theme concerns adoption and engagement. These issues are often intertwined with the previously mentioned limitations of existing AI techniques.

\vspace{4px}\noindent {\bf \colorbox{green!15}{C4}~Integrations with real-world applications.}
While most of the curated papers focus on specific tasks, many also highlight limitations related to integration with real-world applications beyond initial novelty.
Integrations with workflow within real-world applications have been identified as the first challenge.
For example, despite the effectiveness in the controlled lab study, IntroBot~\cite{Shin2023IntroBot} envisions the integration into existing commercial IM applications such as Slack, Teams, and Discord.
We also identified limitations related to scalability and customization. Despite evaluating innovative AIMIC systems in small-group settings, authors of the curated papers highlighted potential challenges when extending these systems to larger teams, more diverse user populations, and long-term, dynamic use cases.
Similar limitations have also been noted in studies of dyadic interactions, such as \cite{Numan2024SpaceBlender}.
The third limitation concerns contextual adaptation, often stemming from the fact that existing AI techniques struggle to adapt to heterogeneous interpersonal communication contexts, such as different meeting types, organizational cultures, and task-specific requirements. For example, Scott~\etal~\cite{Scott2025WhatDoesSuccessLookLike} emphasize the need to evaluate AI-assisted reflection with participants from diverse linguistic and cultural backgrounds.
The final limitation concerns long-term and longitudinal studies, which focus on assessing the effects of AIMIC over time on users' habits, learning outcomes, social dynamics, and organizational culture.
Many of the surveyed papers, such as \cite{Sun2024MetaWriter, deJong2024, Samrose2021, Yao2026PersonaMail, Li2026VizCrit, Leong2024, Kim2025Generations, Sun2024ReviewFlow, Jiang2026ScaffoldedVulnerability}, explicitly call for longitudinal studies to better understand the long-term use of innovative AIMIC experiences.

\vspace{4px}\noindent{\bf Research and design opportunities.}
Our thematic analysis of the curated papers has emerged with four future research and design opportunities (Figure~\ref{fig::challenges}b).
Some research and design opportunities may be implied by the identified challenges and limitations, as noted by the authors of individual papers.
\colorbox{blue!15}{O\#} is used to index each key research and design opportunity.

\vspace{4px}\noindent {\bf \colorbox{blue!15}{O1}~Longitudinal and ecologically valid studies.}
Our curated papers have pointed out that a significant gap exists in understanding the long-term impact of AIMIC experience in naturalistic, real-world settings.
Many curated papers, including those focusing solely on WoZ studies, have highlighted future research opportunities involving field deployments and longitudinal study designs to better understand the real-world use and affordances of specific AIMIC experiences~\cite{Scott2025WhatDoesSuccessLookLike, Samrose2021, Leong2024, Sun2024ReviewFlow}.

\vspace{4px}\noindent {\bf \colorbox{blue!15}{O2}~Opportunities related to the design and integration of AI and agentic AI pipelines.}
Authors of the curated papers highlighted future research opportunities in developing AI models and pipelines that can more effectively understand and adapt to the complex nuances of diverse interpersonal communication contexts.
Our scoping review identified three key capabilities: reasoning about implicit contexts and disambiguating broader user goals~\cite{Scott2025WhatDoesSuccessLookLike}; adapting behaviors, proactivity, and intervention timing based on complex communication contexts~\cite{Lee2025, Chen2025MeetMap, Liu2025}; and interpreting complex non-verbal cues and social signals, such as gestures, facial expressions, and vocal tone~\cite{Bedmutha2024, Yang2025SocialMind, Panda2022}.

\vspace{4px}\noindent {\bf \colorbox{blue!15}{O3}~AIMIC design for agency and accountability.}
We identified key opportunities related to agency and accountability. 
The first theme emerging from our analysis focuses on human agency and autonomy, emphasizing the design of AIMIC systems that empower communication participants, preserve their agency, and avoid fostering overreliance or undermining critical thinking~\cite{Sun2024MetaWriter, Bai2026PartialParticipation}.
The second research opportunity involves understanding how AIMIC systems may reshape power dynamics in interpersonal communication, as well as designing safeguards to prevent the potential misuse of AI~\cite{Bedmutha2024, Johnson2025}.
The final future research opportunity focuses on developing privacy-aware sensing models for AIMIC systems that aim to incorporate complex communication contexts~\cite{Shin2023IntroBot, Yao2026PersonaMail}.

\vspace{4px}\noindent{\bf \colorbox{blue!15}{O4}~Understanding and shaping complex communication dynamics.}
Grounded in specific interpersonal communication contexts, our analysis showed that authors have identified future research opportunities related to understanding how AI may influence interpersonal relationships, group cohesion, social norms, and the psychological impacts of collaboration.\
Key research opportunities include understanding strategies for building and maintaining user trust in AIMIC systems, particularly when errors occur~\cite{Bedmutha2024, Robertson2022}; investigating how AI can be designed to amplify minority voices, mediate conflicts, and foster more inclusive communication experiences~\cite{Lee2025, Houtti2025ObserveAskIntervene}; and examining the effects of AI interventions on emotional contagion and cognitive load during collaboration~\cite{Valente2022EmpathicAuRea, Rayan2025}.

\begin{tcolorbox}[enhanced, colback=darkblue!10, colframe=white, boxrule=0pt, arc=3pt, fuzzy shadow={3.5pt}{-3.5pt}{0pt}{0.4pt}{black!20}]

\textcolor{darkblue}{\textbf{Key Takeaways:}}
Our analysis shows four recurring challenge themes, \incl~generalizability, limitations of enabling AI techniques, disruptions to communication dynamics, and difficulty integrating into real-world workflows. 
Four corresponding opportunity areas emerge in response: longitudinal and ecologically valid studies, more adaptive and context-aware AI pipelines, designs that preserve user agency and accountability, and deeper investigation of AI's effects on group dynamics. 
Many opportunities directly mirror specific challenges, such as latency and hallucination motivating more robust AI pipelines, suggesting a shared if unresolved research agenda. 

\end{tcolorbox}
\section{Discussion}\label{sec::discussion}

\subsection{Research Implication}\label{sec::discussion::implication}

Our analysis characterizes AIMIC along four perspectives: the forms of AIMIC studied (RQ1), how AI is integrated and how human-AI interaction is designed (RQ2), the outcomes and benefits these systems target (RQ3), and the challenges and opportunities the field has surfaced (RQ4). 
Our findings point to broader shifts in how HCI researchers conceive of AI's role in interpersonal communication. 
Our implications are organized into four areas.

\vspace{4px}\noindent{\bf A shift in AI's role as a passive channel to an active communication participant.}
Despite the growth in publication volume (Figure~\ref{fig::trend}), one clearest trend unveiled in our analysis is a qualitative shift in what AI is asked to do. 
Pre-2023 systems predominantly relied on rule-based agents and classifier pipelines to extract, surface, or lightly reformat information that a human still authored and sent (RQ2). 
Since ChatGPT's release, LLM-based systems increasingly initiate action themselves; our analysis along the dimension of initiator/message-sender found that $33$ of the papers in our corpus position AI, rather than a human, as the initiator of AIMIC, with $10$ of these having AI both decide when to act and deliver the message directly to communication partners. 
This mirrors Hancock~\etal's~\cite{Hancock2020} definition of AIMC as an agent that modifies, augments, or generates messages ``on behalf of'' a communicator.
Our analysis further shows that the field moving toward the more autonomous end of that spectrum faster than existing theory has been asked to accommodate. 
Classic CMC research often treated computational systems as a channel through which humans communicate; our analysis shows AI increasingly acting as what Sundar \etal~\cite{Sundar2000} would call a source in its own right, which decides what to say, to whom, and when. 
This reframing matters for design, as the interaction techniques built for a channel (\eg~affordances for editing or dismissing a suggestion) do not automatically transfer to a system that behaves as a third conversational party with its own initiative.

\vspace{+4px}
\noindent{\bf The CSCW matrix explains where AIMIC happens, not how it behaves.}
We adopted the long-standing CSCW time/space matrix \cite{Johansen2020, Rodden1991} as an entry point and two dimensions of analysis.
Our analysis confirms that AIMIC spans the full matrix, with a concentration in synchronous, distributed settings such as videoconferencing.
However, our analysis suggests that time and space alone may not be sufficient to capture the full research landscape of AIMIC.
Two systems can occupy the same cell of the matrix (\eg~synchronous and co-located) and differ enormously in the risks they pose, depending on who initiates AI's involvement, whether its output is private or shared among participants, and how deeply it is embodied in the interaction. 
This is consistent with Hancock~\etal's call~\cite{Hancock2020} for dimensions such as agency and role orientation beyond the CSCW matrix.
We view initiator, message sender, shareability, and embodiment (Figure~\ref{fig::type}, RQ2) not as incidental coding categories, but as fundamental dimensions that should sit alongside time and space, together enabling a richer understanding of how AI behaves when integrated into interpersonal communication.

\vspace{4px}\noindent{\bf Reported benefits may have potential risks.}
Our analysis unveiled that many benefits we identified has a corresponding risk surfaced elsewhere in the corpus. 
Systems designed to provide \insitu~information support risk the cognitive overload and ``productivity paradox'' (\eg~\cite{Bai2026PartialParticipation, Park2024CoExplorer}), in which the effort of attending to AI output offsets the effort it was meant to save. 
Systems designed to intervene and facilitate discussion risk overreliance and eroding user agency. Scott~\etal~found that participants gravitated toward AI-generated answers rather than engaging in the reflection the system was designed to promote. 
Ryan~\etal~\cite{Rayan2025} and Zhou~\etal~\cite{Zhou2026ChatMuse} observed that AI-generated informational support can become a distraction in itself when it fails to align with user needs.
Even reflection-oriented designs, intended explicitly to preserve human judgment, must contend with the calibrated-trust problem~\cite{Lee2004, Shneiderman2020}.
We read this less as a flaw in any individual system than as evidence of a structural tension in AIMIC itself. 
Design choices we found effective at managing this tension includes offering graduated levels of AI involvement rather than a single fixed behavior (\eg~\cite{Chen2025MeetMap}), or assigning AI a visibly partisan role such as a devil's advocate rather than a neutral authority (\eg~\cite{Chiang2024}).
This suggests that future research should examine how well these systems manage the tension between benefits and risks, rather than merely acknowledging the latter.

\vspace{4px}\noindent{\bf Implications of designing future AIMIC experiences.}
We suggested three practical implications for researchers and practitioners designing the next generation of AIMIC experiences, beyond the specific opportunities already outlined in Section~\ref{sec::results::challenges}.
\emph{First}, initiation and shareability should be treated as critical design decisions, since they determine who can be held accountable when AI mediation goes wrong. 
\emph{Second}, because every mode of assistance we surveyed carries a corresponding risk, designers should build in mechanisms for users to observe and adjust how much license an AI mediator has. 
This could mean offering multiple levels of involvement, exposing the AI's reasoning, or making its contributions visually or structurally distinct from human-authored content. Rather than treating a single fixed level of automation as the goal, the aim should be to keep humans meaningfully in control.
\emph{Third}, as AIMIC systems increasingly enter workplaces and specific settings such as classrooms and clinics, evaluation should extend beyond the initial novelty period. 
Most studies in our corpus focuses on generalizable interactive experiences, partly owing to our emphasis on HCI venues over domain-specific ones (Section~\ref{sec::method::datacollection}).
Future scoping reviews may explore how existing AIMIC experiences and techniques translate into domain-specific applications - examining how they are adopted, resisted, or repurposed in those contexts.

\subsection{Limitation}\label{sec::discussion::limitation}

Our scoping review presents a comprehensive understanding of the current design taxonomy of AIMIC systems. However, there are some limitations of our work that are discussed in the following.

\vspace{4px}\noindent{\bf Scope and survey method limitations.}
While curating prior literature using the PRISMA framework, we relied exclusively on OpenAlex~\cite{openAlex} as the source for paper identification.
As our review focuses on AIMIC research within the field of HCI, we limited our corpus to full-paper publications from ACM, IEEE, and Taylor \& Francis.
We selected these publishers because the majority of top-tier HCI conferences and journals are published through these publishers, based on rankings from Google Scholar\footnote{HCI journals by Google Scholar: \href{https://scholar.google.com/citations?view_op=top_venues&hl=da&vq=eng_humancomputerinteraction}{https://scholar.google.com/citations?view\_op=top \\ \_venues\&hl=da\&vq=eng\_humancomputerinteraction}. Accessed on May 16, 2026.} and the CORE conference database\footnote{CORE conference ranking database: \href{https://portal.core.edu.au/conf-ranks}{https://portal.core.edu.au/conf-ranks}. Accessed on May 16, 2026.}.
Nevertheless, we acknowledge that relevant publications may also exist in other venues or under different publishers and were therefore not included in our review.
Future research could expand the scope of paper identification, screening, and analysis to develop a broader understanding of AIMIC beyond HCI.

\vspace{4px}\noindent{\bf More diverse types of research.}
While our review primarily focused on full-paper publications to ensure access to complete and rigorously evaluated research, we acknowledge that relevant AIMIC work may also appear in other formats, such as conceptual and roadmap papers (\eg~\cite{Seymour2024, Wolfe2025}), extended abstracts and work-in-progress (\eg~\cite{Zhou2026ChatMuseEA, Chen2026Proscenium, Brenna2024}), speculative and needfinding papers~(\eg~\cite{Reitmaier2022, Jang2024}), research whose findings and/or prototype may be generalized or provide implications for AIMIC, even though AIMIC is not the primary focus (\eg~\cite{Chen2021ExGSense}), and open-source projects.
We also acknowledge that some critical peer-reviewed publications may have appeared between our data collection cutoff and the writing of this manuscript (\eg~ChatMuse \cite{Zhou2026ChatMuse} and RemiAssist~\cite{Xu2026RemiAssist}, both forthcoming at ACM UIST 2026).
Future studies, therefore, can expand the survey scope to include a broader and more diverse range of research outputs.

\vspace{4px}\noindent{\bf Reported challenges and research opportunities.}
The themes of challenges and research opportunities identified in this review are grounded in those reported by the authors of the curated papers. However, we acknowledge that certain challenges and opportunities may be underreported or omitted altogether. This may stem from factors such as the specific scope of individual studies, manuscript length constraints, or a tendency to overemphasize the strengths and novelty of proposed AIMIC experiences. 
Second, we do not distinguish between challenges for which solutions have already been proposed and those that remain unresolved.
Third, we acknowledge that some challenges and research opportunities may have been investigated in other related fields and research communities. A comprehensive review of this body of work is beyond the scope of this paper. Instead, we see our contribution as an explicit formulation of research opportunities in the context of AIMIC.
We envision this scoping review as an entry point for future researchers and practitioners to understand and design new AIMIC experiences. Readers are encouraged to consult the original papers for a more comprehensive understanding of the contexts, limitations, and design considerations associated with each challenge and opportunity.

\section{Conclusion}\label{sec::conclusion}
This survey presents an in-depth scoping review and understanding of the HCI design taxonomy of AIMIC experiences.
Grounded in the PRISMA approach, we have curated $52$ full-paper publications spanning a range of interpersonal communication contexts and analyzed them in terms of the types of AIMIC, AI integration approaches and human–AI interaction design, reported outcomes and benefits, as well as key challenges and future research opportunities. 
Our findings reveal a research landscape focused on synchronous and distributed communication, with recent work increasingly shifting from traditional AI techniques toward LLM-powered and more agentic AI systems of mediation. 
Across these systems, AI is most often used to provide \insitu~information support, facilitate communication, and augment existing communication workflows, while persistent challenges remain around contextual adaptation, trust and agency, communication dynamics, and integration into real-world settings.
Through this review, we develop a design taxonomy that organizes the emerging AIMIC landscape and highlights opportunities for future AI-mediated designs that preserve human agency and accountability. 
We believe this taxonomy provides a useful foundation for HCI researchers and practitioners to understand, evaluate, and design future AIMIC experiences.

%TC:ignore
\bibliographystyle{ACM-Reference-Format}
\bibliography{reference}

\clearpage
\appendix
\section{All Papers Included in the Review}\label{sec::app::papers}

This section includes all papers in our scoping review.
Table~\ref{tab::app::literature_review} describes the list of papers in our curated corpus. Table~\ref{tab::app::venues} lists the full name of each venue.

{\onecolumn
\begin{longtable}{ >{\centering\arraybackslash}p{0.5cm} >{\centering\arraybackslash}p{1cm} >{\centering\arraybackslash}p{1.5cm} >{\centering\arraybackslash}p{1.5cm} p{9cm}}

\caption{Summary of the curated publications, ordered in reverse chronological order.} \label{tab::app::literature_review} \\

\toprule
\textbf{Year} & \textbf{Citation} & \textbf{Venue} & \textbf{Type} & \textbf{Title} \\ \midrule
\endfirsthead

\multicolumn{5}{c}%
{{\tablename\ \thetable{} -- continued from previous page}} \\
\toprule
\textbf{Year} & \textbf{Citation} & \textbf{Venue} & \textbf{Type} & \textbf{Title} \\ \midrule
\endhead

\midrule \multicolumn{5}{r}{{Continued on next page}} \\ \bottomrule
\endfoot

\bottomrule
\endlastfoot

2017 & \cite{Shi2017IdeaWall} & CSCW & Journal & IdeaWall: Improving Creative Collaboration Through Combinatorial Visual Stimuli \\

2018 & \cite{Andolina2018} & DIS & Conference & Investigating Proactive Search Support in Conversations \\

2018 & \cite{Samrose2018a} & IMWUT & Journal & CoCo: Collaboration Coach for Understanding Team Dynamics during Video Conferencing \\

2019 & \cite{Chandrasegaran2019TalkTraces} & CHI & Conference & TalkTraces: Real-time capture and visualization of verbal content in meetings\\

2020 & \cite{Kim2020} & CHI & Conference & Bot in the Bunch: Facilitating Group Chat Discussion by Improving Efficiency and Participation with a Chatbot\\

2020 & \cite{Aseniero2020MeetCues} & VIS & Journal & MeetCues: Supporting online meetings experience\\

2021 & \cite{McDonnell2021} & CSCW & Journal & Social, Environmental, and Technical: Factors at Play in the Current Use and Future Design of Small-Group Captioning\\

2021 & \cite{Samrose2021} & CHI & Conference & MeetingCoach: An Intelligent Dashboard for Supporting Effective \& Inclusive Meetings\\

2022 & \cite{Liao2022RealityTalk} & UIST & Conference & RealityTalk: Real-Time Speech-Driven Augmented Presentation for AR Live Storytelling\\

2022 & \cite{Valente2022EmpathicAuRea} & VR & Conference & Empathic Aurea: Exploring the Effects of an Augmented Reality Cue for Emotional Sharing Across Three Face-to-face Tasks\\

2022 & \cite{Seita2022} & CHI & Conference & Remotely Co-Designing Features for Communication Applications using Automatic Captioning with Deaf and Hearing Pairs\\

2022 & \cite{Bagmar2022} & GROUP & Journal & Analyzing the Effectiveness of an Extensible Virtual Moderator\\

2022 & \cite{Shin2022} & UIST & Conference & Chatbots Facilitating Consensus-Building in Asynchronous Co-Design\\

2022 & \cite{Panda2022} & CHIWORK & Conference & AllTogether: Effect of Avatars in Mixed-Modality Conferencing Environments\\

2022 & \cite{Robertson2022} & FAccT & Conference & Understanding and Being Understood: User Strategies for Identifying and Recovering From Mistranslations in Machine Translation-Mediated Chat\\

2023 & \cite{Shin2023IntroBot} & CHI & Conference & IntroBot: Exploring the Use of Chatbot-assisted Familiarization in Online Collaborative Groups \\

2023 & \cite{Do2023} & CSCW & Journal & Inform, Explain, or Control: Techniques to Adjust End-User Performance Expectations for a Conversational Agent Facilitating Group Chat Discussions \\

2024 & \cite{Park2024CoExplorer} & DIS & Conference & The CoExplorer Technology Probe: A generative AI-powered Adaptive Interface to Support Intentionality in Planning and Running Video Meetings \\

2024 & \cite{Leong2024} & CSCW & Journal & Dittos: Personalized, Embodied Agents That Participate in Meetings When You Are Unavailable\\

2024 & \cite{Rajaram2024BlendScape} & UIST & Conference & BlendScape: Enabling End-User Customization of Video-Conferencing Environments through Generative AI\\

2024 & \cite{Numan2024SpaceBlender} & UIST & Conference & SpaceBlender: Creating Context-Rich Collaborative Spaces Through Generative 3D Scene\\

2024 & \cite{Chiang2024} & IUI & Conference & Enhancing AI-Assisted Group Decision Making through LLM-Powered Devil's Advocate \\

2024 & \cite{deJong2024} & MUM & Conference & Assessing Cognitive and Social Awareness among Group Members in AI-assisted Collaboration\\

2024 & \cite{Chen2024MemoVis} & TOCHI & Journal & MemoVis: A GenAI-Powered Tool for Creating Companion Reference Images for 3D Design Feedback\\

2024 & \cite{Sun2024MetaWriter} & CSCW & Journal & MetaWriter: Exploring the Potential and Perils of AI Writing Support in Scientific Peer Review\\

2024 & \cite{Bedmutha2024} & CHI & Conference & ConverSense: An Automated Approach to Assess Patient-Provider Interactions using Social Signals\\

2024 & \cite{Sun2024ReviewFlow} & IUI & Conference & ReviewFlow: Intelligent Scaffolding to Support Academic Peer Reviewing\\

2024 & \cite{Rayan2024} & CC & Conference & Exploring the Potential for Generative AI-based Conversational Cues for Real-Time Collaborative Ideation\\

2025 & \cite{Yang2025SocialMind} & IMWUT & Journal & SocialMind: LLM-based Proactive AR Social Assistive
System with Human-like Perception for In-situ Live Interactions\\

2025 & \cite{Zhang2025Understood} & UIST & Conference & Understood: Real-Time Communication Support for Adults with ADHD Using Mixed Reality\\

2025 & \cite{Zhang2025WSCoach} & IMWUT & Journal & WSCoach: Wearable Real-time Auditory Feedback for Reducing Unwanted Words in Daily Communication\\

2025 & \cite{Liu2025} & CHI & Conference & Proactive Conversational Agents with Inner Thoughts\\

2025 & \cite{Johnson2025} & CSCW & Journal & Exploring Collaborative GenAI Agents in Synchronous Group Settings: Eliciting Team Perceptions and Design Considerations for the Future of Work\\

2025 & \cite{Asthana2025} & CSCW & Journal & Summaries, Highlights, and Action Items: Design, Implementation and Evaluation of an LLM-powered Meeting Recap System\\

2025 & \cite{Chen2025AreWeOnTrack} & CHI & Conference & Are We On Track? AI-Assisted Active and Passive Goal Reflection During Meetings\\

2025 & \cite{Houde2025} & IUI & Conference & Controlling AI Agent Participation in Group Conversations: A Human-Centered Approach\\

2025 & \cite{Hu2025} & UIST & Conference & Thing2Reality: Enabling Spontaneous Creation of 3D Objects from 2D Content using Generative AI in XR Meetings\\

2025 & \cite{Rayan2025} & CI & Conference & Cueing the Crowd: LLM-Driven Conversational Cues Across Different Meeting Modalities Increase Topical Diversity of Generated Ideas\\

2025 & \cite{Scott2025WhatDoesSuccessLookLike} & CHIWORK & Conference & What Does Success Look Like? Catalyzing Meeting Intentionality with AI-Assisted Prospective Reflection\\

2025 & \cite{Ma2025} & CSCW & Journal & Nods of Agreement: Webcam-Driven Avatars Improve Meeting Outcomes and Avatar Satisfaction Over Audio-Driven or Static Avatars in All-Avatar Work Videoconferencing\\

2025 & \cite{Chen2025MeetMap} & CSCW & Journal & MeetMap: Real-Time Collaborative Dialogue Mapping with LLMs in Online Meetings\\

2025 & \cite{Kim2025Generations} & CHI & Journal & Bridging Generations using AI-Supported Co-Creative Activities\\

2025 & \cite{Houtti2025ObserveAskIntervene} & CHI & Conference & Observe, Ask, Intervene: Designing AI Agents for More Inclusive Meetings\\

2025 & \cite{Li2025} & WebSci & Conference & Emails by LLMs: A Comparison of Language in AI-Generated and Human-Written Emails\\

2025 & \cite{Scott2025WhatDoesSuccessLookLike} & CHIWORK & Conference & What Does Success Look Like? Catalyzing Meeting Intentionality with AI-Assisted Prospective Reflection\\

2026 & \cite{Johnson2026} & CHI & Conference & ``I Felt Bad After We Ignored Her'': Understanding How Interface-Driven Social Prominence Shapes Group Discussions with GenAI\\

2026 & \cite{Wang2026SeeSawBot} & CHI & Conference & SeeSawBot: An LLM-Driven Chatbot Mediating Across Private and Shared Slack Channels to Support Team Dynamics\\

2026 & \cite{Bai2026PartialParticipation} & CHI & Conference & Enabling Partial Participation in Remote Meetings\\

2026 & \cite{Gunasekaran2026CLARA} & TOCHI & Journal & CLARA: AI-Mediated Facilitation for Enhancing Group Cognition and Cohesion in Remote Collaboration\\

2026 & \cite{Yao2026PersonaMail} & IUI & Conference & PersonaMail: Learning and Adapting Personal Communication Preferences for Context-Aware Email Writing\\

2026 & \cite{Jiang2026ScaffoldedVulnerability} & CHI & Conference & Scaffolded Vulnerability: Chatbot-Mediated Reciprocal Self-Disclosure and Need-Supportive Interaction in Couples\\

2026 & \cite{Li2026VizCrit} & CHI & Conference & VizCrit: Exploring Strategies for Displaying Computational Feedback in a Visual Design Tool\\

\end{longtable}
\twocolumn}

{\onecolumn
\begin{longtable}{>{\centering\arraybackslash}p{2cm} p{12cm}}

\caption{Acronym of the venues, ordered alphabetically by acronym.} \label{tab::app::venues} \\

\toprule
\textbf{Acronym} & \textbf{Venue} \\ \midrule
\endfirsthead

\multicolumn{2}{c}%
{{\tablename\ \thetable{} -- continued from previous page}} \\
\toprule
\textbf{Acronym} & \textbf{Venue} \\ \midrule
\endhead

\midrule \multicolumn{2}{r}{{Continued on next page}} \\ \bottomrule
\endfoot

\bottomrule
\endlastfoot

CC & ACM Creativity and Cognition Conference \\

CHI & ACM CHI Conference on Human Factors in Computing Systems \\

CHIWORK & ACM Symposium on Human-Computer Interaction for Work \\

CSCW & ACM SIGCHI Conference on Computer-Supported Cooperative Work \& Social Computing \\

DIS & ACM Conference on Designing Interactive Systems \\

FAccT & ACM Conference on Fairness, Accountability, and Transparency \\

GROUP & ACM Conference on Supporting Group Work/Sociotechnical Studies \\

IMWUT & Proceedings of the ACM on Interactive, Mobile, Wearable and Ubiquitous Technologies \\

IUI & ACM Conference on Intelligent User Interfaces \\

MUM & ACM International Conference on Mobile and Ubiquitous Multimedia \\

TOCHI & ACM Transactions on Computer-Human Interaction \\

UIST & ACM Symposium on User Interface Software and Technology \\

VIS & IEEE International Conference on Visualization \\

VR & IEEE Vitual Reality Conference \\

WebSci & ACM Web Science Conference

\end{longtable}
\twocolumn}

%TC:endignore

\end{document}